\documentclass[aps, pra, a4paper, amsfonts, amssymb, amsmath, reprint, showkeys, nofootinbib, twoside, superscriptaddress]{revtex4-1}

\usepackage[utf8]{inputenc}
\usepackage{amsmath}
\usepackage{amssymb}
\usepackage{graphicx}
\usepackage{xr}
\usepackage{hyperref}
\newcommand*{\addFileDependency}[1]{
\typeout{(#1)}
\@addtofilelist{#1}

\IfFileExists{#1}{}{\typeout{No file #1.}}
}\makeatother

\usepackage{xcolor}

\newcommand{\eq}[1]{Eq.~(\ref{#1})}

\begin{document}

\title{Unifying the description of hydrocarbons and hydrogenated carbon materials with a
chemically reactive machine learning interatomic potential}

\author{Rina Ibragimova}
\email{rina.ibragimova@aalto.fi}
\affiliation{Department of Chemistry and Materials Science, Aalto University,
02150 Espoo, Finland}
\author{Mikhail S. Kuklin}
\affiliation{Department of Electrical Engineering and Automation, Aalto University,
02150 Espoo Finland}
\author{Tigany Zarrouk}
\affiliation{Department of Chemistry and Materials Science, Aalto University,
02150 Espoo, Finland}
\author{Miguel A. Caro}
\affiliation{Department of Chemistry and Materials Science, Aalto University,
02150 Espoo, Finland}

\date{12.09.2024} 

\begin{abstract}
We present a general-purpose machine learning (ML) interatomic potential for
carbon and hydrogen which is capable of simulating various materials and molecules
composed of these elements. This ML interatomic potential is trained using the Gaussian
approximation potential (GAP) framework and an extensive dataset of C-H configurations
obtained from density functional theory. The dataset is constructed through iterative training and structure-search techniques that generate
a broad range of configurations to comprehensively sample the potential energy surface.
Furthermore, the dataset is supplemented with relevant bulk, molecular, and high-pressure
structures. Finally, long-range van der Waals interactions are added as a locally parametrized
model. The accuracy and generality of the potential are validated through the analysis of
different simulations under a wide range of conditions, including weak interactions, high
temperature, and high pressure. We show that our CH GAP model describes different problems such
as the formation of simple and complex alkanes, aromatic hydrocarbons, hydrogenated amorphous
carbon (a-C:H), and CH systems at extreme conditions,
while retaining good accuracy for pure carbon materials. We use this model to generate
hydrocarbons of different sizes and complexity without prior knowledge of organic chemistry rules,
and to highlight intrinsic limitations to the simultaneous description on intra and intermolecular
interactions within a single computational framework.
Our general-purpose ML interatomic potential has the capability to
significantly advance research in the field of H-containing carbon materials and compounds,
particularly in the areas where longer dynamics, reactivity and large-scale effects may be important.
\end{abstract}

\maketitle

\section{Introduction}

Organic compounds comprising carbon and hydrogen (CH) are among the building blocks of life and are fundamental to biochemistry, medicine, environmental science, and countless industrial applications such as, for instance, the pharmaceutical, petrochemical, plastics, and textile industries~\cite{Chu2012, pr8030355, ARAVAMUDHAN201467, IQBAL2019417, maugeri2012oil}.
Particularly, hydrocarbons and their derivatives are the most important chemicals for energy production and storage, as well as widely used petrochemicals for other industrial purposes.
These compounds display remarkable versatility due to carbon's unique chemical bonding capabilities, resulting in a vast array of complex structures and functional groups. 
The ability to form long chains, branched structures, and cyclic compounds allows carbon to construct a variety of molecules with different shapes and sizes, and bulk and amorphous solids with different properties.
The presence of multiple bond types, such as double and triple bonds, in organic CH compounds can lead to enhanced reactivity, allowing them to participate in various chemical reactions.
Furthermore, CH compounds often exhibit isomerism, where two or more molecules have the same molecular formula but different structural arrangements. 
Understanding the microscopic properties of these materials and compounds is extremely important for establishing a link between fabrication conditions, structure, properties and, ultimately, application-specific performance. 
Eventually, the ability to simulate the chemical reactions and atomistic mechanisms that take place during the industrial processing of C- and H-containing substances can improve their prospects to replace current industry standards based on non-renewable sources and achieve sustainable ways of producing fuels and chemicals.
Simultaneously, the formation of hydrocarbons and related substances plays a crucial role in understanding the formation of planets, comets and other astronomical objects, as well as in establishing the origin of life \cite{Kwok2016}.

Empirically fitted force fields, commonly used in biochemistry, can handle conformational changes and the dynamics
of organic molecules in the absence of chemical reactions. However, the chemical complexity and structural
transformations described above makes theoretical simulations of organic compounds unfeasible within these
computationally affordable empirical methods whenever breaking or formation of chemical bonds are involved.
On the other hand, quantum-mechanical methods like the popular density functional theory (DFT) provide the best tradeoff between accuracy and computational cost, while accounting for chemical reactions.
However, DFT is unable to simulate compounds with more than
a few thousands of atoms per simulation cell, or over extended times, because of the high computational cost. 
The cells required for the simulation of complex CH compounds, whether large molecular chains or hydrogenated
amorphous carbon (a-C:H) materials, are usually much bigger than in typical simulations of inorganic compounds
carried out at the DFT level. 

Thus, there is a clear need for developing a cost-effective model capable of providing a unified description
of hydrocarbons and hydrogenated carbon materials to understand the relations and transformations between the
two within a wide range of conditions.
The general description of such systems has become feasible with the development of data-driven approaches
in atomistic modeling. 
The ability of machine learning (ML) tools to fit high-dimensional data has enabled accurate description of
atomic interactions in materials with low computational cost~\cite{deringer_2019, caro_2023, Gabor_2017, Tkachenko_2021, Deringer_review2021}.
Thus, ML interatomic potentials (MLPs) allow us to tackle problems which are beyond the reach of quantum-mechanical or empirical methods.
By virtue of their computational advantage, MLPs allow us to accurately represent the atomic interactions throughout longer length and time scales than is currently possible with DFT.

In this paper, we introduce a general-purpose MLP for C and H based on the
Gaussian approximation potential (GAP) framework \cite{bartok_2013}, a robust MLP approach
based on kernel ridge regression. Trained from a dataset of different configurations
representing as many parts of the potential energy surface (PES) as possible, our
CH GAP can flexibly simulate CH-based
systems, unifying the methodological framework needed for the description of CH-containing materials
\textit{and} molecules. First, we present the methodology for training general-purpose MLPs with
emphasis on constructing a relevant dataset. Starting with a baseline potential based on small
CH-containing molecules,
taken from the QM9 database~\cite{ruddigkeit2012,ramakrishnan2014quantum},
we perform iterative training and structure search to generate many configurations to cover large areas
of configuration space. Later, the database is complemented by more relevant bulk, molecular,
and high-pressure structures. Then, we add the long-range vdW interactions as a locally
parametrized model~\cite{muhli_2021b} trained with molecules from QM9, melt-quenched structures, and other selected molecular networks.
Finally, we showcase the predictive power of the resulting potential by realizing different
computational experiments on various hydrocarbons, hydrogen-doped carbon materials and a-C:H.
Both the GAP~\cite{ibagimova_2024_GAP} and training database~\cite{ibagimova_2024_dataset} are publicly available for further use by the community.

\section{Methodology}

\subsection{GAP architecture}

We employed the GAP framework~\cite{Bartok2010, Bartok_tutorial}
to train a reactive MLP for C and H. GAP is a kernel-based approach for obtaining ML force fields.
In this framework, the reference PES is sampled via a series of DFT calculations. From these, the
GAP potential learns a \textit{local} approximation to the given PES, allowing us to make
predictions for much larger systems based on the similarity between atomic descriptors of test and
training configurations. We will not describe the approach in full as a detailed account has been
given by Bart\'ok and Cs\'anyi~\cite{Bartok_tutorial} and in the more recent review by Deringer
\textit{et al}.~\cite{Deringer_review2021}. We will focus on the system-specific key
ingredients of the model instead.

In GAP, we regress the potential energy from the fitting coefficients, obtained during
training, and kernel functions, derived on the fly. GAP predictions are made by comparing the
atomic descriptor of a current (computed) structure to a subset of the structures in the database
(the ``sparse'' set). The measure of similarity for each comparison is called the kernel, bounded
between 0 (the two structures are completely different) and 1 (the structures are identical up to
symmetry operations). The predicted energy is, then, expressed as an atom-wise sum of the
products of kernels and fitting coefficients for different types of descriptors,
\begin{align}
\overline{\epsilon}_i = e_{0,i} & + (\delta^{(2b)})^2  \sum_{s} \alpha_s^{(2b)} k^{(2b)}(i,s) \nonumber
\\ & + (\delta^{(3b)})^2  \sum_{s} \alpha_s^{(3b)} k^{(3b)}(i,s) \nonumber
\\ & + (\delta^{(mb)})^2  \sum_{s} \alpha_s^{(mb)} k^{(mb)}(i,s) \nonumber
\\ & + \text{core} + \text{vdW},
\label{eq:01}
\end{align}
where $\overline{\epsilon}_i$ is the predicted local atomic energy, $\alpha_s$ is the machine-learned fitting
coefficient, $k(i,s)$ is the kernel between the atomic environment $i$ and the different atomic
environments $s$ in the sparse set, $e_{0,i}$ is the species-specific energy offset or constant energy
per atom, and $\delta$ is a parameter that controls the energy scale of the model when more than
one descriptor type (2b, 3b, and mb, explained below) is present in the fit. The core contributions to the
atomic energy accounts for the very strong (``exchange'') repulsive interaction at very short interatomic
distances, and are tabulated from the isolated pair-wise interaction curves (H and C dimers, as well as
the CH ``molecule''). The vdW terms are included within a vdW- or dispersion-correction scheme. Both core
and vdW contributions are explained in more detail below.

The choice of a suitable combination of descriptors is crucial to achieve an accurate and data-efficient
approximation of the PES. Our CH GAP combines two-body (2b), three-body (3b) and many-body (mb)
atomic descriptors, and the potential energy of the system is consequently decomposed into individual
contributions, each stemming from an individual descriptor. The atom-wise assignment of these descriptor
contributions can be easily done attending to whether an atom participates in a descriptor (2b; assign half
of the contribution) or is at its center (3b and mb; assign the whole contribution).
In \eq{eq:01}, the first summation term contains the 2b descriptor and the second term stands for the
3b descriptor. The 2b descriptors were incorporated using a 5~\AA{} cutoff radius for the C--C, C--H,
and H--H interactions, whereas 3b descriptors are computed using a 2~\AA{} cutoff for the possible
triplets. The triplets are defined as a central atom with two neighbors within the cutoff distance
from the central atom (these two neighbors are not necessarily within the cutoff distance from each other).
They consist of the six possible permutations of C and H: C--\textbf{C}--C, C--\textbf{C}--H,
H--\textbf{C}--H, C--\textbf{H}--C, C--\textbf{H}--H, and H--\textbf{H}--H, with the bold symbol
indicating the central atom. The inclusion of 2b and 3b descriptors increases the stability of
the GAP and allows us to build a solid baseline for more complex interactions. 

While 2b and 3b interactions are crucial to stabilize the fit, the most important contribution to achieve
quantitative agreement with DFT is the last sum, which is over mb descriptors.
Smooth atom-centered atomic density representations within a given cutoff are used in many of the state-of-the-art
MLPs~\cite{willatt_2019}. In GAP, the smooth overlap of atomic positions (SOAP) many-body
descriptors~\cite{bartok_2013} are used most commonly. We use a more data-efficient and accurate
modification of the SOAP descriptor~\cite{caro_2019}. Note that, rigorously, SOAP and related descriptors,
as usually implemented, are not truly many-body descriptors but, rather, are equivalent to an ensemble of
3b descriptors~\cite{musil_2021}. Their practicality stems from the fact that they can be evaluated much
more efficiently than the corresponding ensemble of 3b terms. For explicit interactions beyond 3b, the
atomic cluster expansion (ACE) provides a systematic recipe for constructing body-order expansions for
atom-centered neighborhoods up to arbitrarily high order~\cite{drautz_2019}. Within the terminology
of the \texttt{gap\_fit} code used to train our GAP~\cite{klawohn_2023}, the used 2b, 3b and mb
descriptors are referred to as \texttt{distance\_2b}, \texttt{angle\_3b} and \texttt{soap\_turbo}. The
full recipe is available within the metadata of our released CH GAP~\cite{ibagimova_2024_GAP}.

Tabulated ``core'' potentials are added for the 2b C--C, C--H and H--H interactions, explicitly
describing the highly repulsive regime when two atoms are within 1~\AA{} from each other. The
explicit inclusion of this term improves the stability and accuracy of the GAP fit significantly,
as well as making high-pressure, high-temperature and collision/deposition (up to a couple hundred of eV)
simulations possible \cite{Caro_spC, Caro_depos2020}. Mathematically, the tabulated per-pair core interaction (together with the $e_{0,i}$
per-atom
energy) is removed from the DFT reference values before the ML fitting coefficients are derived, making
the target PES significantly smoother, since the pair interactions will tend to zero (rather than the physical
Coulomb-like singularity) as two atoms get arbitrarily close to one another. Since the GAP is by construction
smooth, the task of fitting this smoother PES is easier than fitting the true PES. This short-range pair
interaction will be added back at prediction time. In both cases (removal and addition), core potentials
are interpolated using cubic splines from the tabulated values. For reference, the core potentials we use
are plotted in the Supporting Information (SI).

\subsection{Building the dataset}

\begin{figure}[t]
    \centering
    \includegraphics[width=\columnwidth]{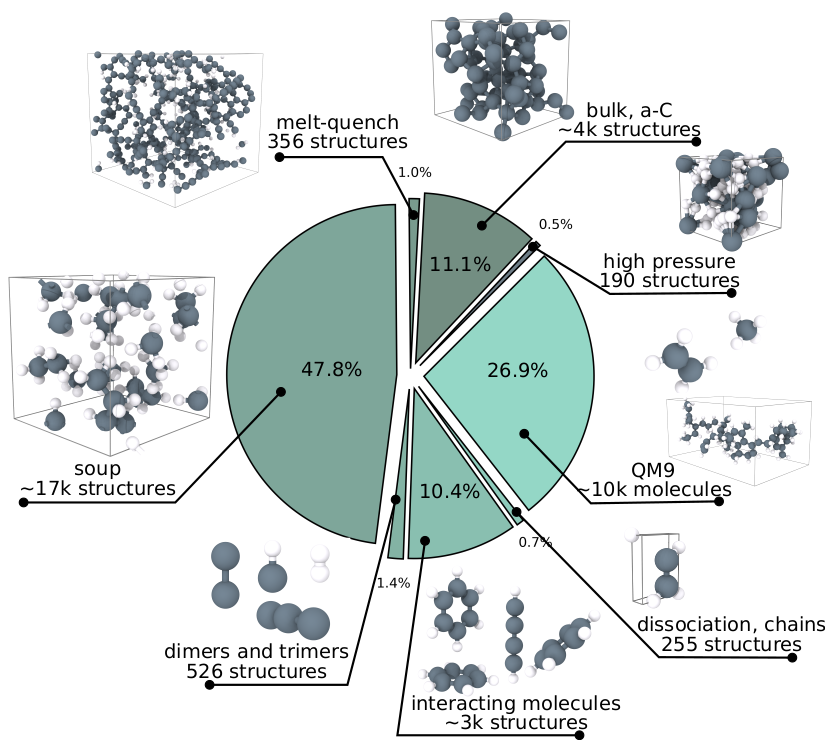}
    \caption{Overview of the structures in the database used for training the CH GAP.}
    \label{fig:overview}
\end{figure}

\begin{figure*}[t]
    \centering
    \includegraphics[width=\textwidth]{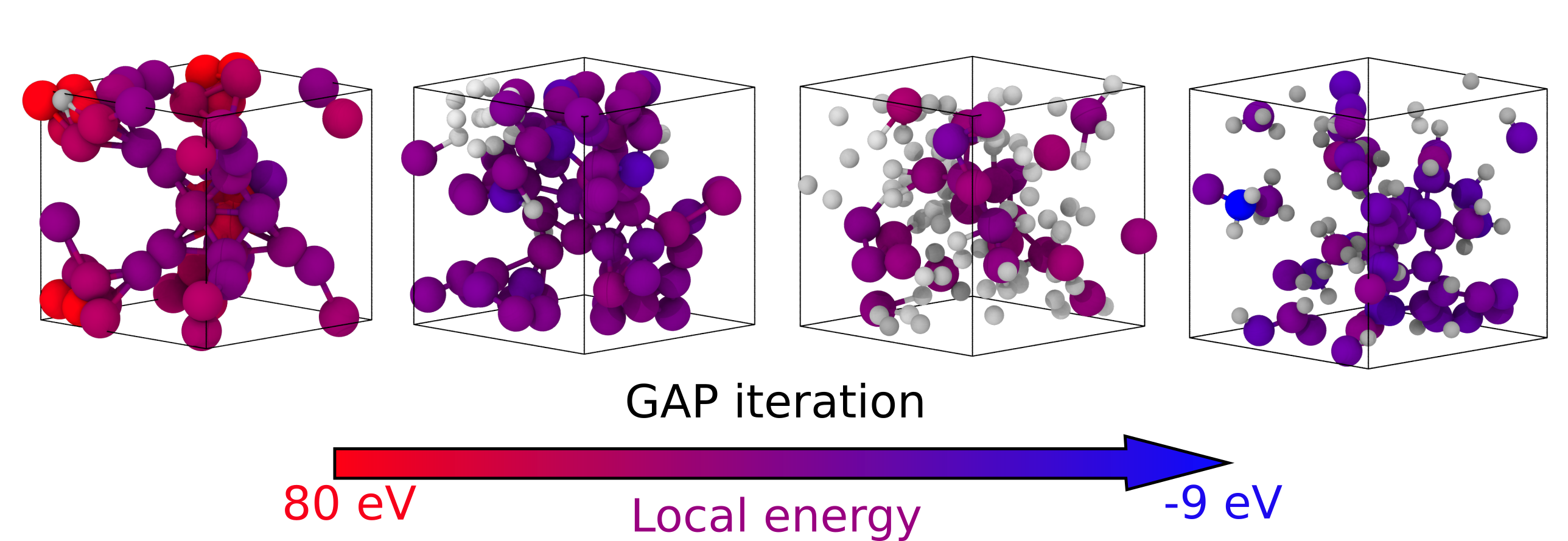}
    \caption{Evolution of the predicted structures during the iterative training of the CH ``soup'' GAP. Initial
    iterations of the GAP model produce structures that are high in energy with unphysical chemical
    connectivity. Later iterations organically learn how to assemble stable hydrocarbons.}
    \label{fig:soup}
\end{figure*}

An MLP relies on two basic ingredients: a methodological framework, described in the previous section,
and a database of training configurations. A general-purpose potential must especially include a
considerable number of configurations. An overview of the configurations used in our CH GAP is shown
in Fig.~\ref{fig:overview}. The whole database was computed at the DFT level of theory with the PBE
exchange-correlation functional~\cite{Perdew1996GeneralizedGA} using
VASP~\cite{Kresse1996EfficientIS,Kresse1999FromUP}. Note that all the individual calculations in the
database use exactly the same convergence parameters to avoid introducing additional noise in the
data. Single-point (SP) DFT energies were obtained using a cutoff energy of 650~eV and automatic
\textbf{k}-point grid generation with the minimum allowed spacing between \textbf{k} points of 0.25 \AA$^{{-1}}$. The automatic \textbf{k}-point grid generation is used in order to accommodate different cell sizes. All molecular dynamics (MD) simulations were carried out with the TurboGAP code~\cite{turbogap,PhysRevB.100.024112}.

\subsection{Molecules}

Training an accurate general-purpose MLP requires large amounts of data. Fortunately, most
low-energy structures are derived by fulfilling different chemical bonding rules, and to a first
approximation they resemble a collection of characteristic atomic fragments or motifs. Thus,
we start by utilizing already available data such as various organic molecules in the QM9 dataset~\cite{ruddigkeit2012,ramakrishnan2014quantum}. QM9 is a comprehensive database of small organic molecules containing C, H, O, N, and F atoms. The whole database contains 134k
molecules. However, we are interested in molecules containing solely C and H. Therefore, we separated
these molecules, resulting in 4898 ($\sim$5k) molecules. Moreover, we randomly distorted the
structures in this subset of QM9 with displacements in each direction up to 0.1~\AA{}. Then, the
total dataset of original and distorted QM9 structures ($\sim$10k structures) was recalculated at the
$\Gamma$-point using DFT with the parameters mentioned above. Adding the QM9 molecules results in the GAP being able to predict energies and forces close to the ground state, whereas adding the distorted structures
allows for sampling the PES further away from the ground state (non-zero forces acting on the atoms).
Therefore, combining these data helps building a much more general and transferable GAP.

\subsection{CH ``soup'' structures}
Next, we employ iterative training to explore the PES and provide data in regions of configuration
space corresponding not only to low potential energy but representing the pathways in configuration space that connect high- and low-energy structures~\cite{a_c_Deringer}. 
The iterative training approach involves training multiple versions of the GAP, where the most recent one is employed to make new structures.
A selection of the resulting structures are then recalculated using PBE-DFT and subsequently supplied as training data for the next iteration of GAP training.
This approach allows us to learn from the successes \textit{and} failures of the previous iterations. The iterative process is repeated until the performance reaches a satisfactory level and the accuracy of the GAP is improved.
The following workflow was adopted: an arbitrary mixture of C and H atoms (corresponding to the
stoichiometry of propane at first) was prepared and simulated using \textit{ab-initio} MD starting
at 15000~K for 10~ps. Snapshots of these calculations were then collected throughout the \textit{ab-initio} MD trajectory to establish a starting dataset.
At such high temperatures, the resulting structures consisted of dissociated atoms
distributed in the simulation box in the form of a dense liquid. 
Therefore, we refer to the contents of the simulation box as a CH ``soup''. This initial \textit{ab-initio} soup database was used to fit the first GAP potential.
With the first GAP potential, we ran MD calculations starting from 15000~K and the energies of the resulting MD snapshots are calculated with DFT. Then, the GAP is retrained with the updated database and the new potential
is used to perform MD at 14000~K, and the procedure of retraining the potential with new data is repeated. The sequence of MD simulations and training new
GAPs continues until the simulation temperature reaches 300~K, where at each temperature a new GAP was trained based on the updated database with a subset of the new structures that were computed with single-point DFT.
The temperature intervals were set to 1000~K between runs at 15000~K and 9000~K, and to 500~K between 9000~K and 500~K.
Thus, during this iterative process 25 GAPs were trained.  

Our goal is to not only improve the accuracy of the potential for a single stoichiometry,
but to explore the PES as comprehensively as possible and obtain a database which can serve to fit a
general-purpose CH MLP. 
Thus, we employed ``branching'' of the iterative training. 
Iterative training of each branch starts from a different set of molecules (with different C$_x$H$_y$
stoichiometries) at 15000~K, using the GAP trained on \textit{ab-initio} MD snapshots. 
Overall, we trained 10 branches with different starting configurations, where each branch was trained iteratively over different temperatures.
Using this approach, we have generated the soup database containing 21k structures.
The evolution of the database during the iterative training is shown in Fig.~\ref{fig:soup} with
example configurations. 
The carbon atoms are color-coded according to their local energies, where red is a high local energy and blue is a low local energy. 
For instance, the first generations of the soup structures correspond to very
high energy regions of the PES, where the generated structures are unphysical,
with local energies up to 80~eV/atom. 
However, the CH GAP starts to correctly distinguish between PES regions of high and low energy over the iterations. 
This results in correctly producing structures in more energetically favorable parts of the PES for later GAP
generations. 
For instance, in Fig.~\ref{fig:soup}, the structure in the last panel contains hydrogenated amorphous carbon (a-C:H) with methane and ethane molecules and local energies are already down to $-9$~eV (similar to the cohesive energy per atom of graphite).
Over 25 generations in each branch and 10 branches, GAP learns from completely unreasonable structures, where atoms are clumped together with very short bond lengths, with hydrogen and carbon atoms segregated into different areas of the simulation box, to structures with adequate bonding and stoichiometry. 
The structures in the last iterations show the expected formation of simple alkanes and a-C:H.

\begin{figure}[t]
\centering
\includegraphics[width=0.9\columnwidth]{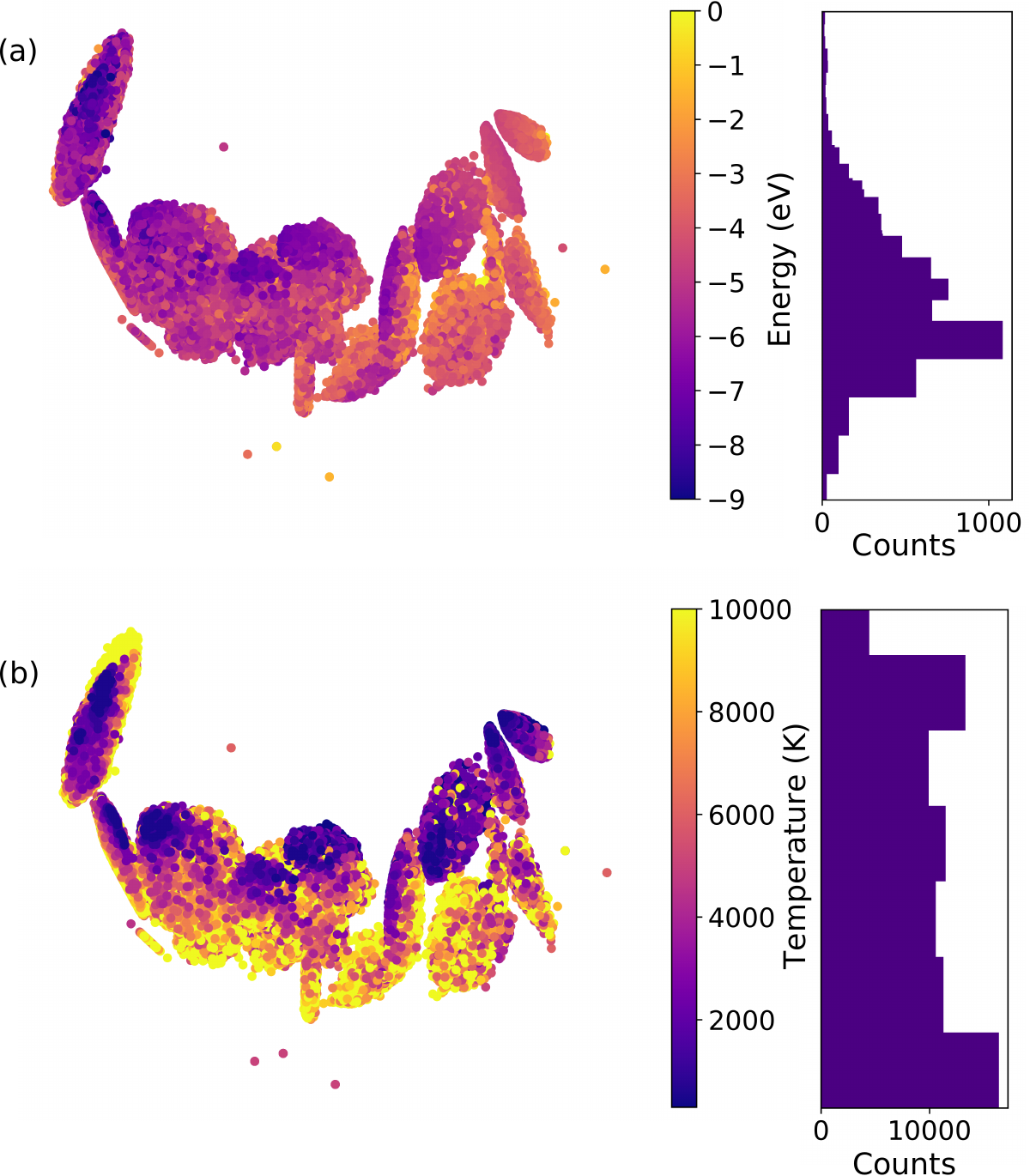}
\caption{2D embedding of a SOAP dissimilarity metric through the cl-MDS algorithm. The map contains only soup structures and the color coding highlights (a) the distribution of the structures by local energy and (b) the distribution of the structures by generation temperature.}
\label{fig:clmds_soup}
\end{figure}

To further analyze the soup structures, we use the k-medoids data-clustering technique to construct a map of similarity of data points~\cite{kmedoids}. 
The provided dataset of structures is partitioned into clusters according to their structural similarity, measured via SOAP kernels like those used in the construction of the GAP MLP~\cite{de_2016,cheng_2020}. The structural motif with a minimal dissimilarity to all other structural motifs in the cluster is called a medoid. 
We used the fast-kmedoids library for the k-medoids computation and the cl-MDS algorithm to graphically represent the clustering via low-dimensional embedding~\cite{hernandezleon2022clusterbased}. 
The structural similarity map for the soup database is shown in Fig. \ref{fig:clmds_soup}, where panel (a) is color coded according to local energies (we restrict the energy values to those below 0~eV) and panel (b) shows the same map but color coded according to the temperatures at which the given structures were generated. 
The similarity map shows how high-temperature (liquid-like) local motifs tend to populate their own high-energy clusters as well as the periphery of other clusters where they coexist with low-energy structures: transiently, structures resembling metastable configurations (especially sp chains) will be created within the liquid. Distorted molecular building blocks (e.g., strained molecules and radicals) will also populate the periphery of some clusters where the central dark areas correspond to the low-temperature stable arrangements: alkanes, alkenes, alkynes, aromatics, a-C motifs, etc. The coexistence in the training database of high- and low-energy structural motifs that are structurally related can be considered a positive sign, as this will allow the GAP to map the regions of configuration space corresponding to structural transformations that are relevant to chemical reaction barriers. While we do not expect achieving ``chemical accuracy'' (i.e., errors in the predicted reaction barriers below 1~kcal/mol) given the general-purpose nature of our GAP, such comprehensive sampling will prevent the MLP from extrapolating in these regions and will ensure the stability of the fit over a wide range of thermodynamic conditions.
Thus, we may conclude that the sampling of the PES is comprehensive and each soup branch in our structure-generation protocol enables access to a part of the PES with different structures and different energies.
         
It is important to mention that, during iterative training, we did not add molecules from
the QM9 database, resulting in GAP learning itself the formation of stable molecules by quenching from
a hot gas precursor. 
This is an outstanding result, showing that, with a systematic sampling of the configuration space and inclusion of a large variety of dynamically generated diverse structures, GAP can learn different parts of the PES and predict the formation of stable structures without any prior knowledge of the system at hand.

\subsection{Dimers, trimers, and condensed phases}
At this point, combining QM9 and soup datasets provides us with a comprehensive database to train a reliable \textit{general-purpose} CH GAP. 
However, we further refine the dataset by adding dimers (H-H, C-C, and C-H) with the distances from 0.4~\AA{} up to 4.5~\AA{} and trimers with the distances from 0.4~\AA{} up to 2.8~\AA{}. 
Incorporating these configurations provides a robust baseline for the potential at very small interatomic separations, as well as represents the simplest cases of bond breaking. Furthermore, we add a ``core'' potential, which is fitted to reproduce the dimer curves down to 0.1~\AA{} separation (for the carbon dimer, this corresponds to 1.8~keV). This core potential is a tabulated pairwise interaction (with cublic spline interpolation in between data points) that is subtracted from the energies and forces in the database before fitting the MLP, and added back at prediction time. This makes the PES that needs to be fitted significantly smoother whenever atoms are very close to one another (e.g., for high-temperature structures), improving the stability and accuracy of the fit significantly. 
Finally, another important addition to the database is the inclusion of condensed phases of pure carbon as a foundation to ensure the predictions remain accurate in carbon-based matter with low (or no) H content. 
To add these configurations, we rely on the comprehensively tested earlier work on construction of carbon GAPs~\cite{a_c_Deringer, Heikki_vdw}, with
amorphous carbon structures, diamond and graphite configurations from Ref.~\cite{a_c_Deringer}, and numerous surface structures from Ref.~\cite{Heikki_vdw}.

\subsection{Intermolecular interactions}
The potential trained using the database with the above-mentioned structures showed good overall performance. 
However, this GAP failed in accurately capturing weak interactions between specific molecules, for instance, the configuration involving two benzene molecules (Fig.~\ref{fig:overview}). 
Addressing this issue poses a non-trivial problem: how to balance intra versus intermolecular interactions. 
First, the intrinsic ability for characterizing the intramolecular interactions is limited by the cutoff radius of the SOAP descriptor. 
For instance, a SOAP with a cutoff of only 4~\AA{} would not be able to capture all the atoms in a benzene molecule, where H atoms are up to 5~\AA{} apart. 
This limitation, which is inherent to all local descriptors used to construct MLPs, can be fixed to some extent by increasing the cutoff radius. However, this will be accompanied by a dramatic increase in both computational cost and the complexity of the configuration space to be represented, because the number of atoms within a cutoff sphere scales as the cube of the cutoff radius. An alternative way to effectively increase the ``cutoff'' is to use message-passing MLPs~\cite{batzner_2022,batatia_2022}, which do nevertheless also suffer from poor scaling as a function of the number of neighbor layers, which in turn determine the effective extended range of the interactions.
Second, the intermolecular interactions cannot be described specifically, since there is no explicit attribution of the atom to a molecule. Thus, neighboring atoms contribute to an atomic density descriptor such as SOAP based solely on how far they are from the central atom, and not whether they belong to the same or a different molecule than the central atom. This is problematic: intra and intermolecular interactions contribute very differently to the total energy of the system (about two orders of magnitude on the energy scale) but these differences are not reflected in the descriptor.

Improving the scarcity of data in the relevant part of configuration space may help to represent these interactions more accurately. 
Therefore, we specifically add a subset of different molecule pairs, that we refer to as interacting molecules. 
The subset is generated using active learning and uncertainty-based configuration selection to improve the GAP's ability to handle complex intermolecular interactions. 
In order to generate the subset, we started with randomly chosen pairs of CH-containing molecules from the QM9 database up to 7 carbon atoms.
The probability of selecting the molecules is set based on the energy and size of the molecule. 
The probability is lower as the energy above the convex hull is higher. 
A bigger size of the molecule also lowers the probability to favor the inclusion of small structures.
Then, we estimate the uncertainties for the new structures based on how far away from the existing interacting molecules in the training set they are (in configuration space)~\cite{Deringer_review2021}, and identify those with the largest expected errors. By stochastically generating many structures of interacting molecules, we can select the ones that are most likely to improve the next trained GAP. This way, we generated and added to the final database about 3k structures. This subset of interacting molecules is published in Zenodo \cite{interacting_molecules}. 

Section~\ref{sec:nonbonded} delves deeper into the intrinsic limitations to accurately capture intra and intermolecular interactions and how this subset of data may improve the representation of non-bonded interactions. 

\subsection{Regularization}
To avoid overfitting, kernel-based regression models use regularization. GAPs use $L_2$ (or Tikhonov)
regularization, where a term quadratic in the fitting coefficients is added to the objective function,
penalizing against large individual fitting coefficients. The weights added to the fitting coefficients
are called regularization parameters. In the GAP formulation, one can add a different regularization
parameter per observable, i.e., one per total DFT energy, three for each force, and six for each stress
tensor~\cite{Bartok_tutorial,Deringer_review2021,klawohn_2023}.
The regularization parameter, often denoted by $\sigma$ and given in the same units as the
target observable, quantifies the expected ``noise'' in the input data; in the case of
GAP, it also quantifies the (in)ability of the model to learn the data, e.g., because of the assumption
of locality. A bigger regularization parameter allows the model to deviate more from the data, whereas
a smaller regularization parameter forces the model to follow the training data more closely.
Considering the diversity of the configurations used in our database (Fig.~\ref{fig:overview}), the
model needs to be regularized differently depending on the type of structures. This is important to
achieve a high accuracy for low-energy structures like stable molecules and bulk crystals, while
maintaining a reasonable description of amorphous and high-energy atomic arrangements. Bearing in
mind the different features of the dataset, we set the values of the regularization parameters as follows.

Intuitively, the configurations representing the ground state or known to be stable over a wide range
of conditions should be regularized to a higher target precision (small $\sigma$). The disordered
configurations require lower accuracy (big $\sigma$). More rigorously, one can think about the adjustment of the regularization parameters in terms of giving more weight in the fit to configurations that contribute more to the system's partition function. The summary for the different regularization schemes is given in the SI (Table S1). Molecular
data from QM9~\cite{ramakrishnan2014quantum} is fitted to high precision while soup structures are fitted to one order of
magnitude lower precision. Moreover, different regularization parameters were used even within the soup
structures depending on the temperature at which they were generated: much smaller regularization
parameters were used for the soup structures generated at 300~K than for the structures generated
at 15000~K. We are implicitly expecting a linear dependence between temperature and energy
fluctuations above the low-temperature ground state.
Therefore, the typical choice for the regularization of the first generation was about 0.025 eV/atom, whereas all subsequent values of $\sigma$ are decreased linearly with
temperature, setting the resgularization to 0.001~eV/atom for the last generations. 

\subsection{Sparsification}

To make the fitting of such a large dataset as ours manageable, we need to optimize the selection
of the representative atomic environments within the sparse set. For this, we use a matrix reconstruction
technique called CUR matrix decomposition~\cite{CUR}, implemented in \texttt{gap\_fit}~\cite{klawohn_2023}.
The set of representative structures is chosen separately for different configurations as well, featuring
some of the configurations more than others. This approach helps to reduce the computational cost of fitting
and evaluation while gaining more control over the tradeoff between accuracy and
computational cost of the resulting potential. 

\subsection{Dispersion corrections}

\begin{figure}[t]
\centering
\includegraphics[width=0.9\columnwidth]{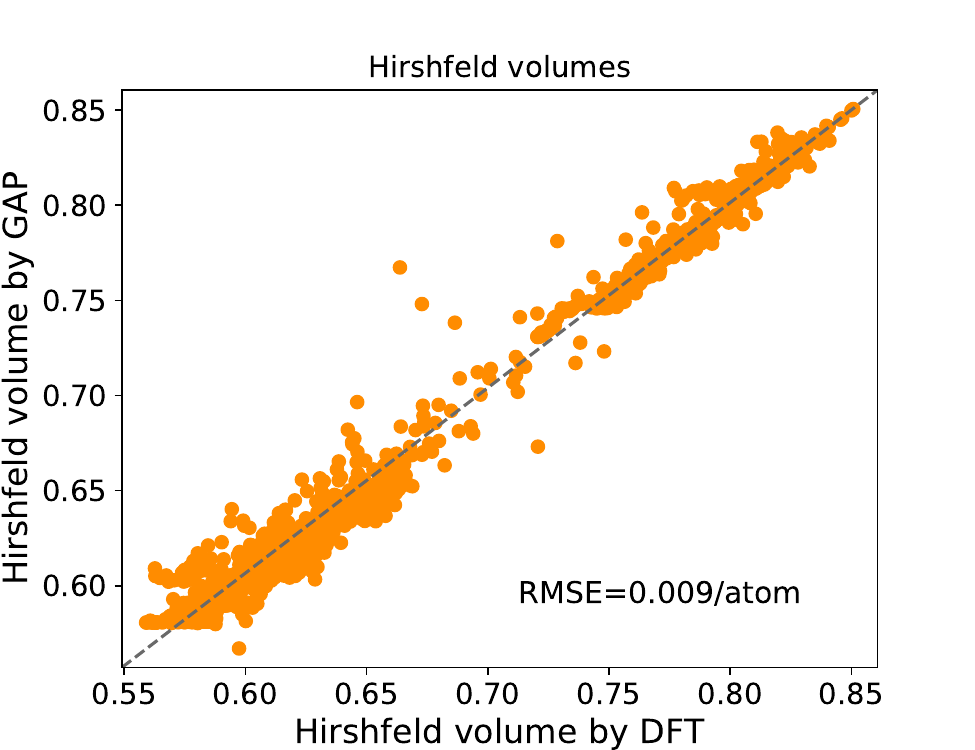}
\caption{Scatter plot of Hirshfeld volumes for C and H calculated for a training dataset of the interacting molecules.}
\label{fig:hirsh}
\end{figure}

\begin{figure}[t]
\centering
\includegraphics[width=0.9\columnwidth]{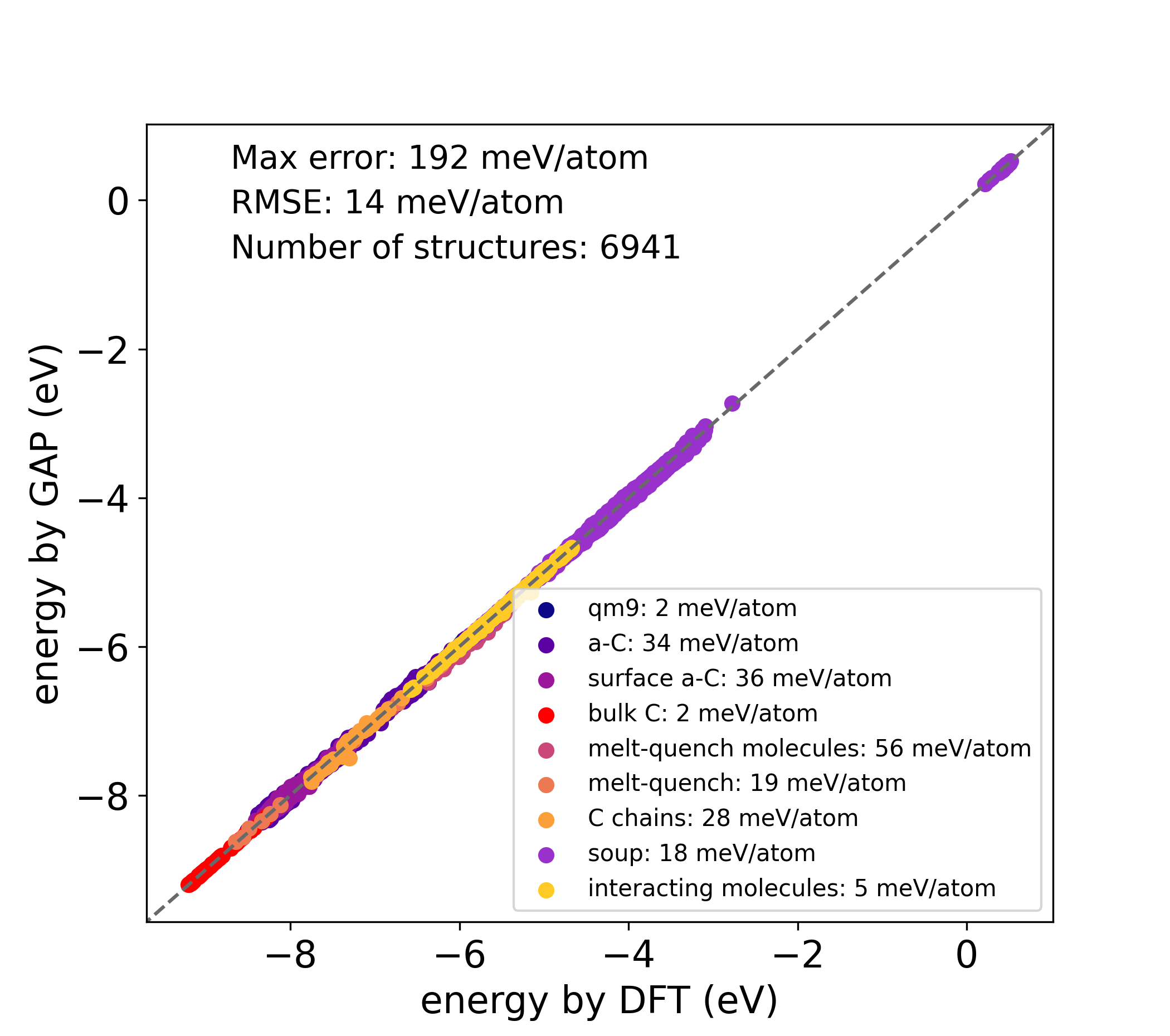}
\caption{Scatterplot for a test set of structures, which were not included in the
ﬁt, comparing DFT calculations with the prediction from GAP. GAP energy \textit{vs.} DFT energy, datapoints color-coded according to the type of structure.}
\label{fig:scatter_test}
\end{figure}

While GAP is able to accurately capture the short-range ``bonded'' interactions,
many target properties of interest in molecular and weakly bonded condensed systems
are strongly influenced by long-range van der Waals (vdW) interactions. Within our
methodological framework, these interactions are incorporated using a local model
of atomic polarizabilities based on Hirshfeld volume partitioning of the charge
density. Hirshfeld volumes can be used to parametrize the relatively simple and
computationally efficient (although somewhat outdated) two-body Tkatchenko-Scheffler
(TS) vdW-correction scheme~\cite{tkatchenko_2009}, as well as the more sophisticated
(and computationally expensive) many-body dispersion (MBD) scheme~\cite{tkatchenko_2012}.
The TurboGAP code supports both TS corrections~\cite{Heikki_vdw} and a linear-scaling implementation of MBD, which became available recently~\cite{muhli2024MBD}.

To train the local model, we calculated the reference Hirshfeld volumes from DFT
using the VASP implementation~\cite{bucko_2013} for the QM9 database and several other
relatively small structures  such as molecular networks from the  soup dataset. 
A local-property model was trained with the \texttt{gap\_fit} code~\cite{klawohn_2023} using the
reference Hirshfeld volumes obtained from these calculations. Then, the local model was
added to the baseline GAP potential. Our results, shown in Fig.~\ref{fig:hirsh},
demonstrate that Hirshfeld volumes can be predicted very accurately from the local
model. In the applications to molecular and materials modeling shown later in this
paper, we use these Hirshfeld volumes together with the TS approach, whenever vdW
corrections are applied. This is equivalent to the way that PBE+TS DFT calculations
are carried out~\cite{tkatchenko_2009}, i.e., the TS correction is applied ``on top''
of the correction-free CH GAP prediction. We note here in passing that other
vdW-correction schemes could also be applied on top of the underlying MLP prediction,
e.g., the D3 method with tabulated vdW parameters~\cite{grimme_2010} in combination
with neural network potentials (NNPs)~\cite{morawietz_2013,ying_2023}.

\section{Benchmarks}

\subsection{General accuracy}

To monitor the overall general accuracy of the potential, we check the CH GAP \textit{vs} DFT energies and forces. 
We split the dataset into test and training sets. 
The energy scatter plot for the test set is shown in Fig.~\ref{fig:scatter_test}. The test set includes 6941 different structures and the energies are color-coded according to the structural configuration type. Although the errors might seem large, with RMSE values of 14~meV/atom for energies, and 1~eV/\AA{} per carbon atom and 0.5~eV/\AA{} per hydrogen atom for forces (see bottom panel of Fig.~S1 (e)-(f)), these are below those reported for a-C~\cite{a_c_Deringer,wang_2022} using the same GAP framework. In fact, the average errors are dominated by the contribution of highly energetic structures: high pressure, high temperature and highly distorted structures, molecular radicals, etc. Thus, it is instructive to split the test set into different types of structures: molecules from the QM9 database, interacting molecules, soup structures, structures from generative melt-quench simulations and their molecular fragments (referred as melt-quench molecules), and pure carbon structures such as carbon chains, amorphous carbon, surface a-C and bulk carbon structures (configurations of graphite and diamond).

Breaking down the errors for each particular structural configuration shows the potential's performance and transferability for different systems.   
The errors for different configurations are depicted on the plot. The lowest errors are observed for the structures with lowest overall energy per atom: 2~meV/atom for QM9 molecules and bulk carbon structures (graphite and diamond), and 5~meV/atom for interacting molecules (whose error is dominated by intermolecular interactions, as we discuss later in detail).
The errors for a-C structures are 34-36~meV/atom, which is expected for the amorphous structures. Overall, the errors for pure carbon structures are comparable to those obtained by MLPs trained solely for pure C~\cite{a_c_Deringer}, which is a good sign of the ability of the CH GAP to retain its accuracy for pure carbon while generalizing for very different CH structures.
Furthermore, the highest error of 56~meV/atom is obtained for the molecules from melt-quench simulations. These were extracted after quenching the CH structures from high temperatures down to 300~K (see the Sec.~\ref{sec:melt-quench}). Many of these structures consist of molecular fragments with strained and broken topologies, radicals, unsaturations and conformers with high energy. Given that these molecules are not well represented in the training dataset, the error is relatively high.

Finally, to verify that our CH GAP is not overfitted, we compare the error computed for the QM9 test set above (2~meV/atom) to the error computed for the whole (training+test) QM9 set.
The scatter plot for CH GAP energies and forces \textit{vs} DFT energies and forces for CH-containing molecules in the QM9 database is given in the top panel of Fig.~S1 (SI). The RMSE for energies is about 3~meV per atom, whereas the errors for the forces are 0.42~eV/\AA{} per carbon atom and 0.16~eV/\AA{} per hydrogen atom. The RMSE for the whole CH-QM9 set is in fact slightly larger than for the test CH-QM9 set, ruling out extrapolative behavior and validating our choice of regularization parameters for the GAP fit (which are listed in the SI). We remark here in passing that our QM9-based dataset includes strongly distorted molecules (necessary to generate a stable MLP) and, therefore, the error is relatively high compared to what could be achievable if training and testing the model on relaxed molecules only, since distorted molecules include very strained topologies that disproportionately contribute to the overall error.

\subsection{Hydrogenation}
\begin{figure}[t]
\centering
\includegraphics[width=\columnwidth]{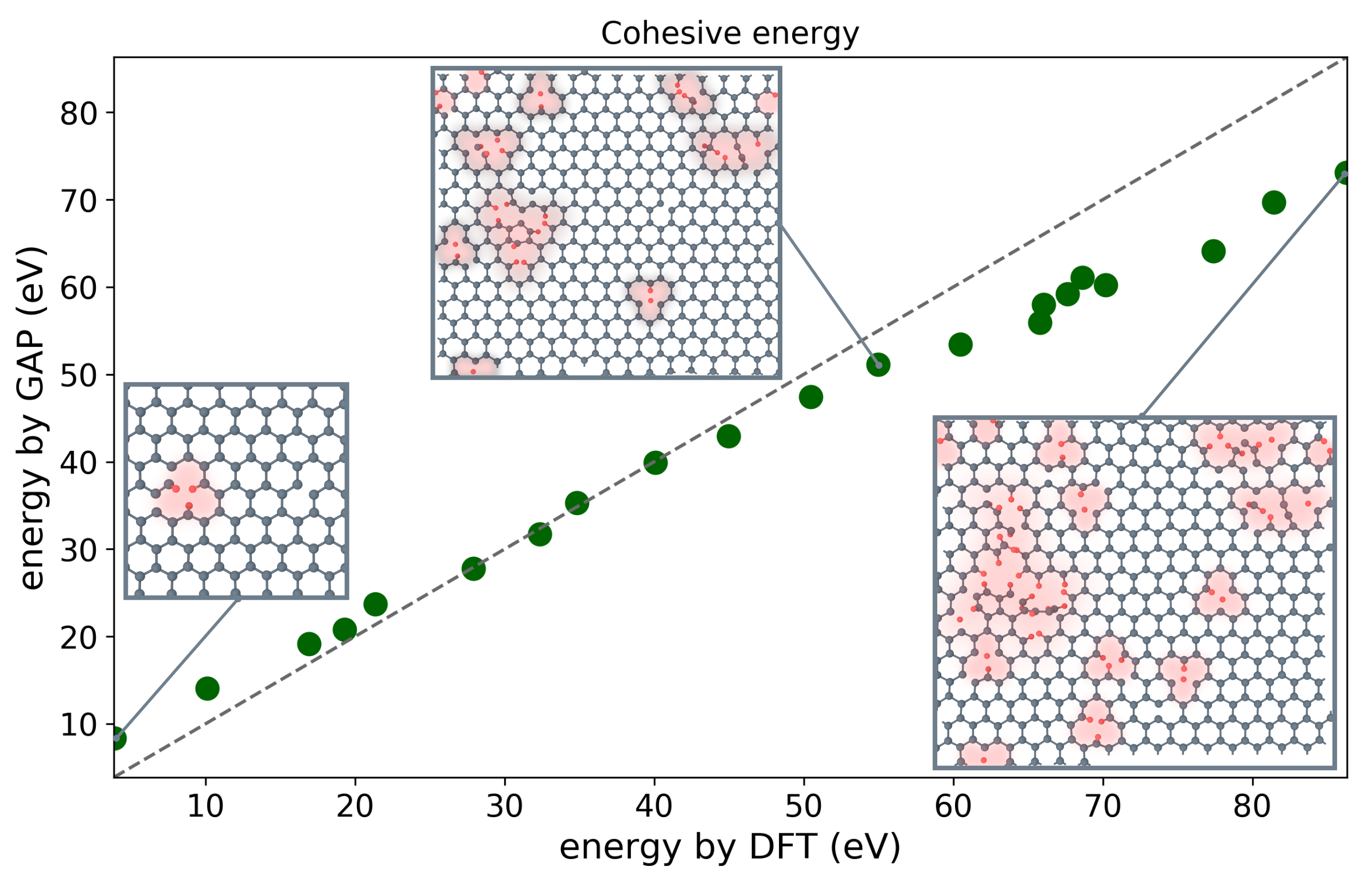}

\caption{Comparison of cohesive energies in defected graphene computed with DFT versus the GAP prediction. The structures contained 512 atoms, and at each step C vacancy
defects were introduced at randomly chosen positions, then 2, 3, or 4 H atoms were placed into a C vacancy. The structures are then relaxed and the DFT and GAP predictions compared. Three example structures are illustrated, where the defective areas are colored in red. The reference energies used in the calculations are the cohesive energy per atom of C in graphene and the same for H in an H$_2$ molecule.}
\label{fig:def}
\end{figure}
Next, we design a test to assess the ability of the potential to predict the formation of arbitrarily complex defects in a graphene sheet.
The defect formation energetics is critical to determine whether the potential can be used in the simulations of the synthesis of hydrogenated carbon materials. 
To address this task, we start with a graphene sheet with 512 atoms, where we create a C vacancy at a random position. Then, we passivate the vacancy with 2 to 4 hydrogen atoms (number randomly chosen). The structure's energy, after GAP relaxation, is then recalculated with DFT. The calculation is performed in a sequential manner, where we used the previous snapshot as starting point for the next simulation, introducing a new defect. The resulting comparison of cohesive energies obtained by GAP \textit{vs} DFT is shown in Fig.~\ref{fig:def}. 
The cohesive energy is very well represented, especially up to a certain amount of created defects. The deviation starts when more and more hydrogen is added to the system and the resulting graphene sheet is much further away from any structures in the training dataset.
Here we note that the training database does \textit{not} contain structures closely resembling these, and thus this is a particularly stringent test of the transferability of our CH GAP to arbitrary problems. If quantitative accuracy for this particular problem was needed, a user could easily retrain the CH GAP (or another MLP flavor) by including these test structures in the training dataset.

\subsection{Non-bonded interactions}\label{sec:nonbonded}

Non-bonded interactions are important for describing the dynamics of molecular systems,
including liquids, gases and molecular crystals. On a first approximation, these interactions
can be separated into long-range electrostatics, long-range vdW,
and short- and mid-range repulsion. Long-range electrostatics, notoriously difficult to capture within MLP
frameworks \cite{Isayev2023, Kulik_2022}, are not particularly relevant in C-H materials
and molecules, where most atomic environments are non polar. \textit{Short-range} electrostatics,
on the other hand, are implicitly captured within the mb descriptor cutoff sphere ``out of the box''. Long-range vdW interactions originate from the dynamical effects in electron
charge distributions. Unlike electrostatics, which can be repulsive or attractive depending on
the sign of the effective atomic charges or relative orientation of the effective molecular
dipoles, vdW interactions are \textit{almost always}~\cite{vanoss_1978} attractive. These interactions are
present in non-polar systems and are in fact crucial to correctly describe the equation of
state of hydrocarbons~\cite{hunter_2013,veit_2019}. Finally, repulsion happens when the
orbitals of different atoms or molecules overlap significantly leading to a high-energy
interaction, such as that prescribed by Pauli's exclusion principle at very short distances
(what we referred to as ``core'' interactions earlier), but also at longer interatomic
distances because of the overlap of the
tails of the orbitals as already predicted within a mean-field PBE-DFT calculation. This
last type of interaction takes place within mid-range interatomic distances and becomes
problematic within MLP formalisms, as we will discuss below.

\begin{figure}[t]
\centering
\includegraphics[]{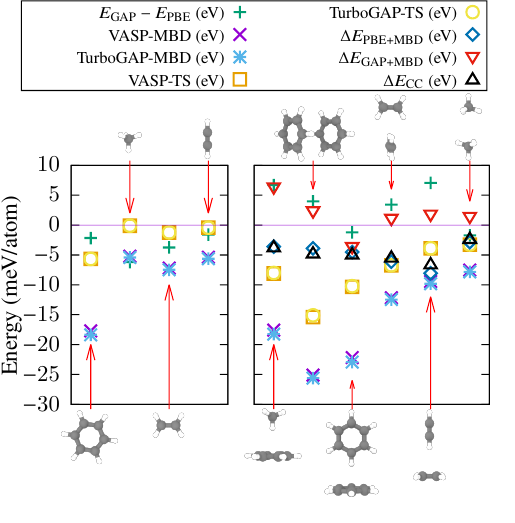}
\caption{Several quantities for different CH-containing molecular pairs from the S22 database. The left panel shows the quantities for the individual molecules and the right panel for the pairs. Shown are the CH GAP error \textit{vs} DFT ($E_\text{GAP}-E_\text{DFT}$), the vdW correction energies according to MBD and TS methods as implemented in VASP and TurboGAP (VASP-MBD, TurboGAP-MBD, VASP-TS and TurboGAP-TS), and the predicted binding energies of the molecular pairs $\Delta E$ as computed with VASP (PBE+MBD), TurboGAP (GAP+MBD) and for the reference CCSD(T) data (CC).}
\label{fig:s22}
\end{figure}

Given the context provided above, in our CH GAP non-bonded interactions are captured via
the implicit learning of PBE-DFT non-bonded contributions to the total energy within the
MLP cutoff (5 \AA{} ), plus the explicit
vdW correction which is applied on top of the baseline CH GAP. The TurboGAP code offers
vdW corrections that rely on Hirshfeld effective volume ML models. Currently, the TS-vdW
correction scheme is implemented in the main code branch and thoroughly tested. The more
sophisticated MBD-vdW correction formalism, including self-consistent screening of atomic
polarizabilities, is fully implemented~\cite{muhli2024MBD} but still not merged into the main code branch. We will use both vdW
correction schemes here to validate our CH GAP and CH Hirshfeld ML model. As a benchmark, we use the set of hydrocarbons included in the S22
database~\cite{jurevcka_2006,takatani_2010} as distributed within ASE~\cite{larsen_2017}.
The results of this test are given in Fig.~\ref{fig:s22}.

The hydrocarbon subset of the S22 database gives binding energies computed at the
coupled-cluster (CCSD(T)) level of theory for two benzene dimers, a methane dimer,
an ethene dimer, an ethene-acetylene pair, and a methane-benzene pair. Our benchmark
compares the PBE-DFT, DFT-TS and DFT-MBD results obtained with VASP to our TurboGAP
predictions based on our ML models. It also compares the binding energies computed with
VASP and TurboGAP with MBD vdW corrections to the CCSD(T) values from
Ref.~\cite{takatani_2010}.
In the left panel of Fig.~\ref{fig:s22} we show the energy predictions for the isolated
molecules. As can be seen, our MBD and TS vdW corrections are essentially undistinguishable
from the VASP reference values. The formation energies predicted by the GAP are within
5~meV/atom of the VASP PBE-DFT values. In terms of establishing the relative stability of
different hydrocarbons, these errors are negligible. However, the situation is different
for the interacting molecule pairs, shown in the right-hand panel. Again, the vdW corrections
are almost identical to the reference VASP values. The errors in learning the PBE-DFT energy
are similar, albeit somewhat larger, than for the isolated molecules, at $\sim 5$~meV/atom.
However, because the \textit{relative} error in the binding energy for the molecule pairs,
which is dominated not by the vdW error but by the GAP error, is significantly larger than
the relative formation energy error, this will lead to sizable quantitative inaccuracies
in determining properties dominated by non-bonded interactions. An example would be the
accurate determination of mass densities at given $(P,T)$ conditions for gas/liquid
hydrocarbons. We note here in passing that these errors will be of the same order of magnitude
as those introduced by using, e.g., TS as the vdW correction scheme. As can be seen from the
graph, DFT-MBD binding energies are very similar to CCSD(T) binding energies, whereas the TS
correction deviates significantly from the MBD correction.

\subsection{Improving the description of intermolecular interactions}

During the course of this work we have extensively (but unsuccessfully) attempted to improve
the quantitative level of description of non-bonded mid-range repulsive interactions. Within
the GAP framework, the main limitation seems to be that the strong bonded interactions and
comparatively weak non-bonded interactions need to be learned with the same structural
descriptor in an electronic-structure agnostic fashion, i.e., without any explicit information
regarding which atoms are bonded and which are not. Because there are about two orders of
magnitude differences between these two energies, trying to improve the description of weakly
interacting molecules (e.g., by tuning down the corresponding regularization parameter) leads
to a worsening of the formation energy prediction of the isolated molecules, and vice
versa \cite{Ioan_npj}.

The most promising strategy that we have identified so far is to effectively separate these
interactions by making the SOAP descriptor able to distinguish between intra and intermolecular
interactions based on structural information alone. Preliminary tests showed that we can
reduce the errors for both intra and intermolecular interactions significantly. E.g., with
non-production-ready models we could reduce these errors from 4.2 to 2.9~meV/atom
(intramolecular energies) and from 6 to 2.6~meV/atom (intermolecular energies)
on representative databases. Our group is currently developing this approach
further and will report on any advances in this regard elsewhere.

\section{Applications}
\subsection{Melt-quench simulations of hydrocarbons and a-C:H}\label{sec:melt-quench}

\begin{figure*}[t]
\centering
\includegraphics[width=0.9\textwidth]{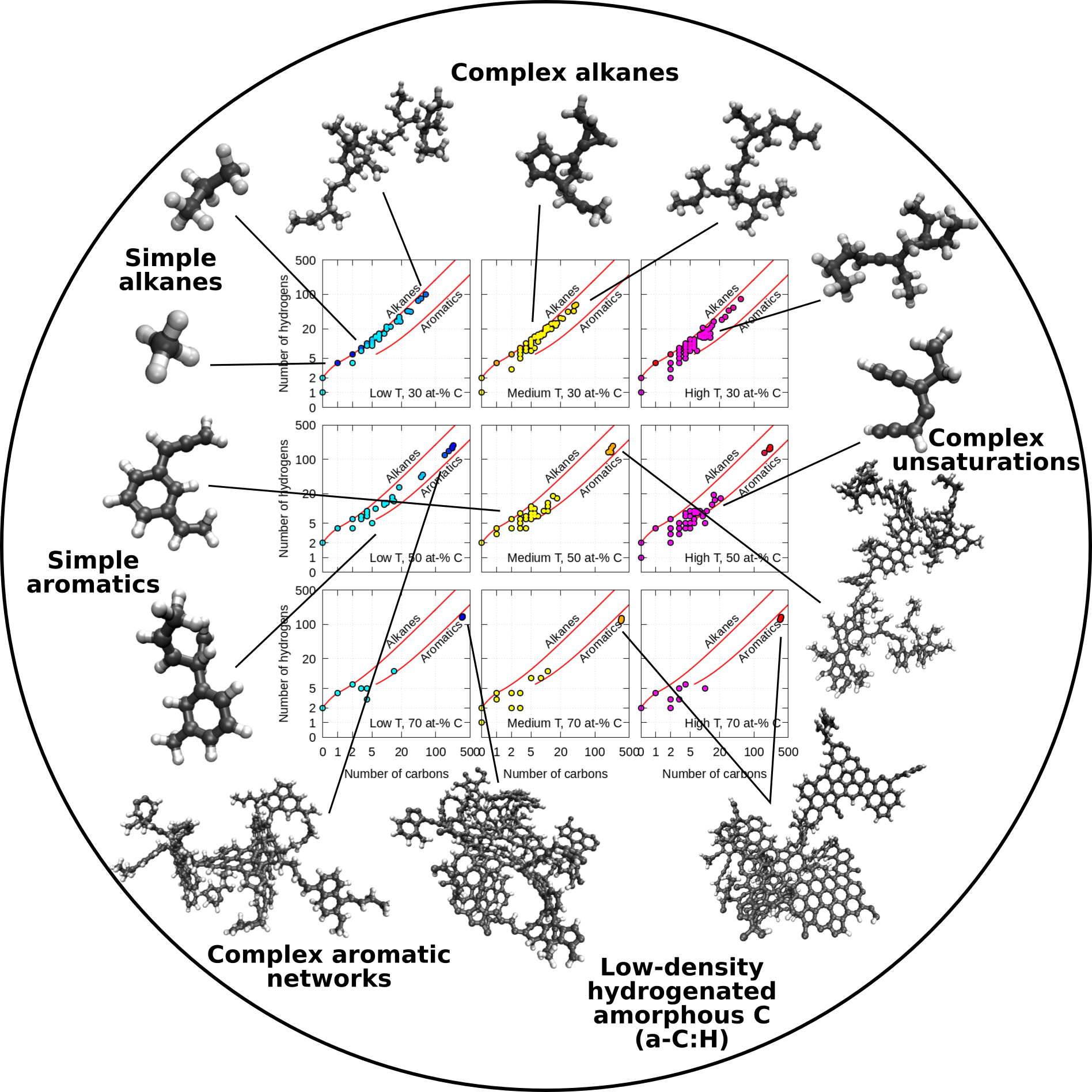}
\caption{Results of melt-quench (``cooking'') simulations at different annealing temperatures
and C:H ratios, showing how the CH GAP is able to generate complex compounds,
from molecules to the solid state. The starting configurations are CH ``soups'' (liquid)
at 5000~K, then quenched to the cooking temperatures of 2500, 3100 and 3700~K (low $T$,
medium $T$, and high $T$ in the figure, respectively), cooked for 100~ps, and finally quenched
to 300~K and 1~bar. Each dot on the different inset panels indicates the stoichiometry of
molecules that were generated as part of the simulation, with the number of C atoms
$n_\text{C}$ given along the horizontal axis and the number of H atoms $n_\text{H}$ given along
the vertical axis. E.g., a dot at coordinates (1,4) indicates the presence of methane.
``Molecules'' with very large $n_\text{C}$ ($> 100$ atoms) are actually a-C:H
solids or solid-like fragments (or nanoflakes) extending across the periodic boundaries of the
simulation box (depicted with ball-and-stick models on the bottom-right corner of the figure).
The red lines are a rough guide to the bonding character, showing the $(n_\text{C}, n_\text{H})$
relations in the ideal alkane ($n_\text{H} = 2 n_\text{C}+2$) and acene (labeled ``atomatics'';
$n_\text{C} = 4n+2, n_\text{H} = 2n+4$) series.}
\label{fig:map}
\end{figure*}

\begin{figure*}[t]
\centering
\includegraphics[width=\textwidth]{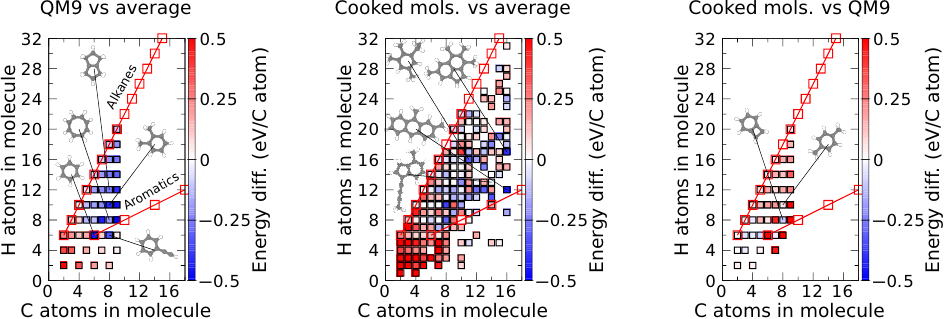}
\caption{Stability analysis of QM9 and cooked molecules. The left panel shows the energy differences between the QM9 molecules and the average energy of the lowest-energy molecules for a given stochiometry, the middle panel shows the energy differences between the molecules obtained during melt-quench simulations and the average energy for the given stochiometry, and the right panel shows the energy differences between the cooked molecules and the molecules from the QM9 database. }
\label{fig:gen_vs_qm9}
\end{figure*}

An efficient (albeit non-comprehensive) way to explore a complex PES is by performing high-temperature
MD. At high temperature, the atoms have enough kinetic energy to overcome local potential energy
barriers and thus probe different metastable configurations. After equilibrating (or ``annealing'')
the system at high temperature, it can be quenched to ``trap'' the atoms in these metastable
arrangements. By varying the density, annealing temperature, annealing and/or quench MD times, and
stoichiometry, one can generate atomistic systems with different arrangements. This melt-quench
approach has been extensively used to generate atomistic models of a-Si and different types of
disordered carbons~\cite{marks_1996,marks_2002,detomas_2016,detomas_2019,a_c_Deringer,caro_2023}.
In this work, we follow the same approach but need to consider different C:H ratios which is obviously
not an issue in elemental compounds. Clear limitations of this approach are, among others: i) it is
not comprehensive nor deterministic as each MD run will generate a different set of products, depending
on random initialization of positions and velocities; ii) it does not necessarily produce the
most thermodynamically stable products, which would require infinite annealing and quenching times;
and iii) it favors high-entropy products, e.g., we found it difficult to generate
benzene following this approach (benzene is low in potential energy but also low in entropy, as a highly
symmetric molecule). More comprehensive methods to explore the PES include random-structure
search (RSS)~\cite{pickard_2011} and nested sampling~\cite{skilling_2004,partay_2021}, both of which have
been used in combination with MLPs with application to carbon~\cite{marchant_2023,karasulu_2022}. Here,
we value the fact that melt-quench simulations allow us to generate highly complex structures for modest
computational cost, which are beyond the reach of RSS and nested sampling.

Therefore, we conducted MD simulations using our CH GAP to investigate the formation of different C- and
H-containing structures ranging from molecules to solids, across different temperature ranges.
Mixtures of C and H, with varying ratios of C and H from 10~at.~\% to 100~at.~\% C were created, where
each structure contains 512 atoms randomly distributed in the simulation cell. Furthermore, in order to
gather statistically significant data, each starting mixture underwent randomization of C and H positions
five times. Then, the structures were equilibrated at 5000~K until they were completely melted. Subsequently,
the molten structures were annealed at different temperatures, from 2500 to 3900~K at 200~K intervals,
from which we report results at 2500, 3100, and 3700~K as most representative. These mixtures were annealed
for 100~ps before being quenched to 300~K at 1~bar pressure. A sample of the different chemical structures
obtained from these simulations as a function of the carbon to hydrogen ratio is shown in Fig.~\ref{fig:map}.
In the figure, the three temperature ranges were categorized as low (2500~K), medium (3100~K), and high (3700~K).   
The degree of aromaticity and, more generally, presence of rings in the derived structures increases with both
temperature and carbon content, as does the number of unsaturations.
Our results demonstrate that the interatomic potential is able to capture extremely complex chemistry
where, depending on the ratio of H to C and quenching rate, very distinct structures can be achieved. For
instance, when the concentration of precursor mixture is 30~at.~\% C, simple alkanes form at 2500~K. At medium and high
temperatures, and for higher C content, the formation of complex alkanes was observed, including an increased
presence of double and triple bonds.
Finally, we obtained low-density hydrogenated amorphous carbon (a-C:H) from mixtures containing 70~at.~\% C.
We plan to study the atomistic structure of a-C:H in more detail in the near future, taking advantage of the
predictive power and computational efficiency of our CH GAP.

We note again the exploratory nature of these simulations. To derive more thermodynamically stable hydrocarbons,
significantly longer MD runs, with longer annealing step and slower quenching, or structure-prediction-specific
algorithms~\cite{pickard_2011,oganov_2006} would be better suited. In addition, the accuracy of the CH GAP could
be improved specifically for high-energy hydrocarbons via iterative training. Here, we give a taste of the kind of
tasks that can be accomplished in this exploratory fashion. Namely, we extracted and examined all the unique
hydrocarbon molecules that were derived from these high-temperature annealing simulations, up to 16 C atoms. A
total of 655 unique molecules were obtained this way. All of these were further relaxed with the GAP and
single-point energies were computed at the PBE-DFT level. With these results, we perform different
comparisons to the QM9 entries.
The results of the analysis are summarized in Fig.~\ref{fig:gen_vs_qm9}.
First, we extract the ground-state molecules for a given stoichiometry, e.g., for the formula
C$_4$H$_{10}$ isobutane is selected over $n$-butane, since it is lower in energy. This selection is done separately
for the QM9 dataset, the ``cooked'' molecules, and the combination of the two.
Second, we perform a linear regression on the total energies of these lowest-energy molecules
for the combined dataset, as a function of the number of C and H atoms. This provides a baseline for the average
total energy of the ground state for a given stoichiometry.
Third, this baseline is subtracted from the energies of the QM9 molecules (left panel of Fig.~\ref{fig:gen_vs_qm9})
and cooked molecules (middle panel) and the difference is plotted. In the figure, red squares denote a higher
than expected energy (i.e., less stable) and blue denotes lower than expected energy (i.e., more stable). We are
already able to note several observations: i) as expected, there is a general trend towards increased stability
per atom as the size of the molecule and its C content increases; ii) also as expected, there are some outliers,
specifically due to ``aromatic stabilization'', e.g.,
benzene, toluene and phenylacetylene; iii) although the QM9 dataset contains no hydrocarbons with an odd number of
H atoms, the cooked dataset has many such entries, often corresponding to (relatively stable) radicals; iv) it is
easy to generate complex, yet stable (in the sense of low in energy), hydrocarbons with the CH GAP,
unlocking the possibility to discover new, potentially synthesizable, organic molecules. Finally, we compare the
cooked molecules (up to 9 C atoms) to the QM9 ones, which is shown in the right panel of the figure. Interestingly,
several molecules are lower in energy that those in the QM9 dataset. We recall that these molecules were never
explicitly included in the training dataset, and that the CH GAP learned how to build them by implicitly inferring
the rules of hydrocarbon chemistry through a fragment-based approach.
We intend to use the CH GAP to predict new hydrocarbons in the near future. In the meantime, we have made the
database of cooked molecules discussed above freely available online~\cite{ibagimova_2024_dataset}.

Thus, our interatomic potential is able to capture complex chemistry involved in the formation of hydrocarbons.
These results are remarkable considering the simple manipulation of the precursor concentration and temperature,
which demonstrates the versatility and predictive capability of our CH GAP. We speculate that this GAP could be used
in diverse applications, for instance to understand the presence and properties of hydrocarbons in molecular
clouds in outer space, where the harsh environment and exposure to high-energy radiation can trigger the formation
of hydrocarbons seldom observed on Earth~\cite{Kwok2004, Barzaga2024, alata2015}.

\subsection{Methane at low temperature}
\begin{figure}[t]
\centering
\includegraphics[width=0.8\columnwidth]{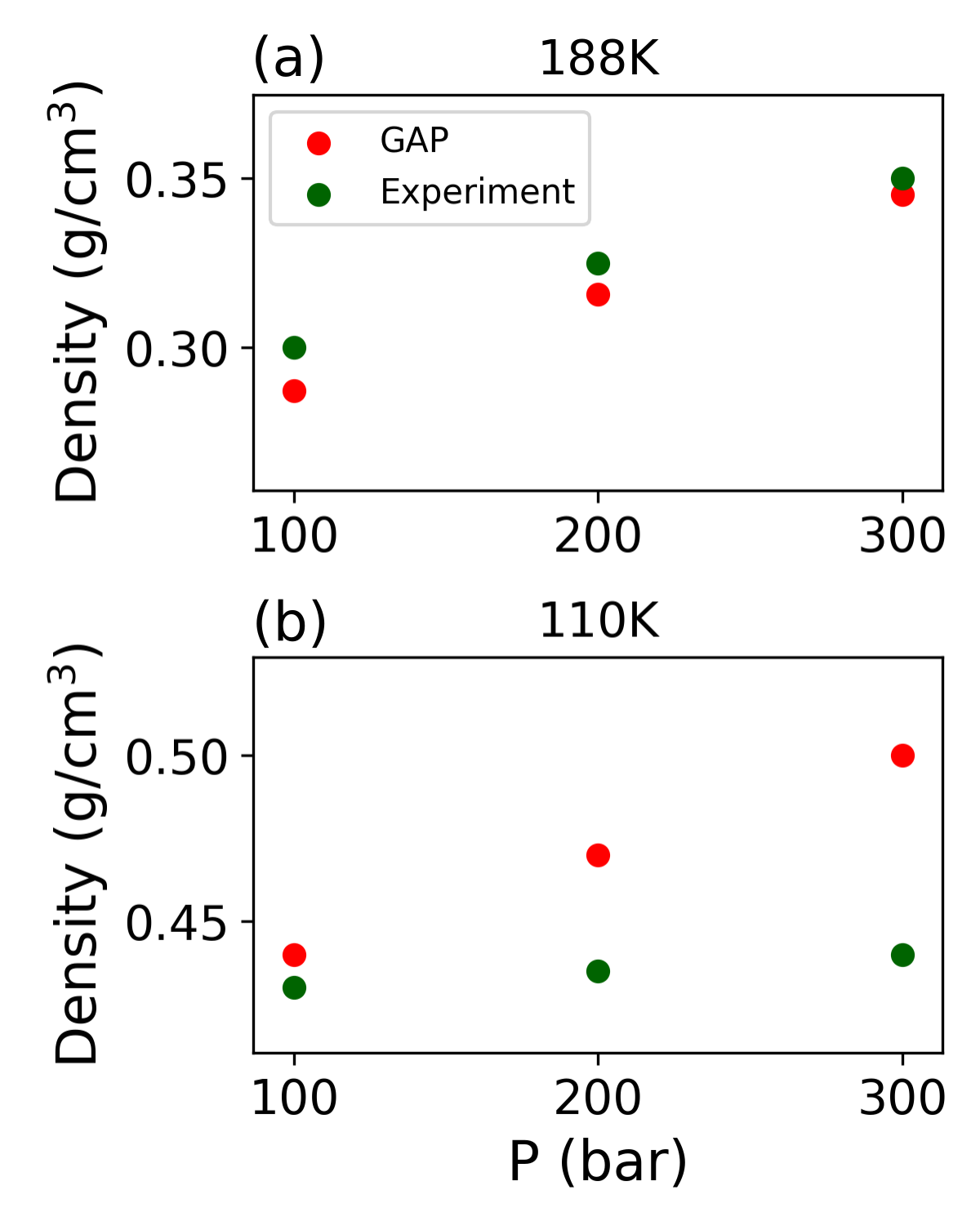}
\caption{\label{fig:eos} Density of methane at (a) 188~K and (b) 110~K. Red dots represent data predicted by GAP and green dots are experimental data from Ref.\cite{Goodwin1972}.}
\end{figure}

To assess the efficacy of the included van der Waals interactions, an important test involves studying methane, particularly at low temperatures. Methane exhibits a complex behavior at low temperatures, undergoing several phase transitions within a narrow temperature range. At room temperature, methane exists in the gas phase, but at temperatures below 200~K, it transforms into a liquid, and at even lower temperatures, below 100~K, it crystallizes. Therefore, by analyzing the behavior of methane at these temperatures and pressures, we gain valuable insights into the performance of van der Waals interactions, which play a significant role in describing the properties of molecular systems like methane under various conditions.

However, we know that TS vdW corrections overestimate the strength of vdW interactions, thus, our vdW ML local model might be affected. To correct for this effect, we tried tuning several parameters in our simulations. First, the standard reference atomic vdW radius $r_0$ is empirically obtained for crystals, thus, we used smaller values for methane: 1.452~\AA{} for C and 1.001~\AA{} for H. Second, another important parameter that has to be tuned, $s_R$, is the parameter of the damping function in the TS vdW correction method. To optimize the value of $s_R$, we performed several calculations on the density of liquid methane by varying this parameter and selected $s_R = 1.15$, which results in the correct density. Using these empirically fitted parameters, the equation of state of liquid methane was calculated at 110~K and 188~K for different pressure values: 100, 200, and 300~bar, as illustrated in Fig. \ref{fig:eos}. The calculations were conducted as follows. A simulation box with 512 randomly placed CH$_4$ molecules (2560 atoms) was created with low density. First, the system was equilibrated at 110 and 188~K for 40~ps and the given pressure. Then, the box was slowly scaled to the desired density over 30~ps. The resulting density was then obtained at every value of the pressure. Our results indicate very good agreement between the CH GAP calculations and experimental data of CH$_4$ density at 188~K. However, at 110~K, there is a growing discrepancy between our computed density and experimental measurements at higher pressure values. This difference may be attributed to the range of various other factors beyond the scope of van der Waals interactions. Particularly noteworthy among these is the role of quantum nuclear contributions~\cite{veit_2019}, which we have disregarded. 

\begin{figure}[t]
\centering
\includegraphics[width=\columnwidth]{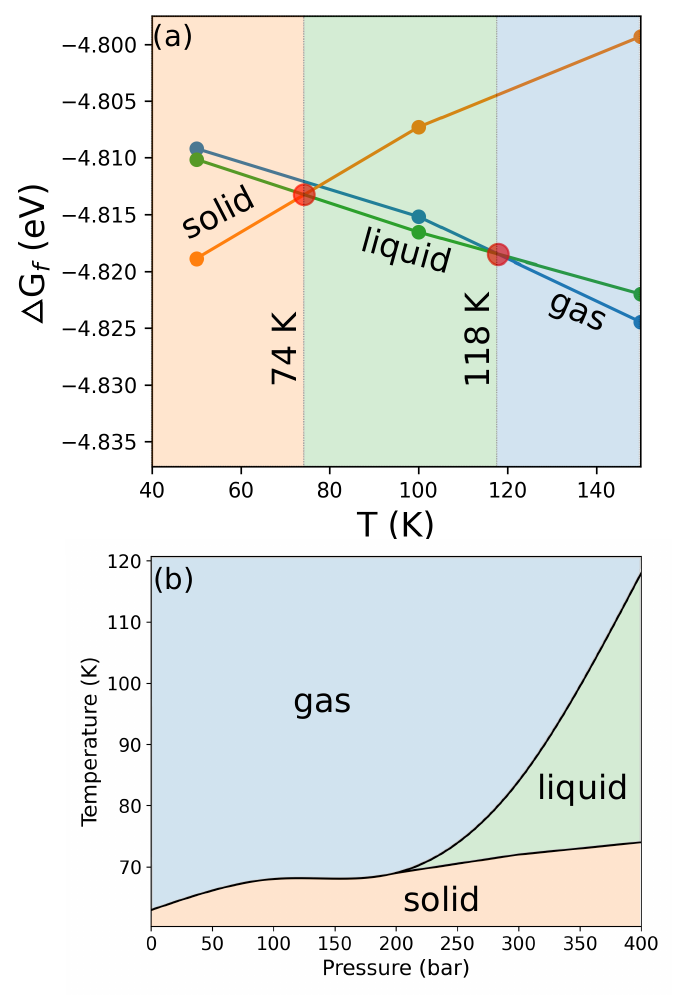}
\caption{\label{fig:phase} (a) Gibbs free energies of formation per methane molecule in the solid, liquid, and gas phases as a function of temperature at the pressure of 400 bar. (b) Phase diagram of methane derived from our GAP simulations.}
\end{figure}

To subject the CH GAP potential to a more challenging evaluation, we simulated the solid-liquid-gas phase diagram of methane. Foremost, we acknowledge that the quantitative prediction of methane's phase diagram is beyond the scope of this study. The phase diagram of methane at low temperatures is extremely complicated due to a coexistence of solid, liquid, and gas phases in a very narrow temperature window and presence of challenging phenomena such as non-trivial many-body vdW interactions and quantum nuclear effects. Furthermore, accurately calculating the phase diagram requires determining the associated free-energy landscape with exceptional precision, as well as requiring a statistically significant amount of sampling. Our objective in this study is to test the limits of the CH GAP potential by using deliberately complex test cases. Therefore, we recommend consulting Refs.~\cite{Partay2016, Ashton2022-el, Bore2023-ek} for various methodologies to sample the free-energy landscape, and Refs.~\cite{veit_2019, Gomaa2024-db, Ranieri2024-tg} for the thermodynamic properties of methane.    
To probe the solid-liquid-gas phase stability, we used the two-phase thermodynamics (2PT) method~\cite{lin_2003} to compute the free energies of the molecular ensemble. This method relies on integrating the density of states, which we derived from MD simulations.
We applied the 2PT method to the gas, liquid and solid phases to directly compare their free energies. By evaluating the free energy at different thermodynamic conditions, we identified the stable phase as the one with the lowest free energy. First, solid, liquid, and gas systems with 540, 540, and 2560 atoms, respectively, were prepared. The gas and liquid systems are distinguished by their density. Then, these systems were equilibrated at 200~K in the case of the gas phase, at 100K in the case of the liquid phase and at 50~K in the case of the solid phase, for 20~ps. Next, the structures were simulated at 1, 100, 200, 300, and 400~bar pressure with the temperature starting from 50~K for solid and liquid, and from 200~K for gas phase for 40~ps. The temperature and pressure were controlled with the Bussi thermostat and Berendsen barostat with a time constant of 100~fs and a time step of 0.5~fs. Then, the box dimensions were averaged for another 20~ps to be assigned for an NVT run in the subsequent step. For the next set of temperatures, the last snapshot from the previous simulation was used, where the system is heated (in the case of the solid and liquid) and cooled (in the case of gas) for 20~ps. That way, we could sample the temperatures increasing/decreasing from opposite directions at 50~K, 100~K, 150~K, and 200~K. At each temperature and pressure, we carried out free-energy calculations using the 2PT method implementation in the DoSPT code \cite{dospt, dospt_redox}.  The Gibbs free energy of the structures at 400 bar as a function of temperature for each phase is shown in Fig.~\ref{fig:phase}(a). The crossing lines indicate the phase transition temperatures, which are 74~K for the liquid-to-solid phase transition and 118~K for the gas-to-liquid phase transition. Using this method, we compute the Gibbs free energies of the three phases as a function of the temperature for each pressure value and approximate the phase coexisting conditions to construct a phase diagram. The full computed phase diagram of methane is depicted in Fig.~\ref{fig:phase}(b).

The stability of the solid phase was determined to be up to 63~K at low pressures and up to 74~K at high pressures. The stability of the liquid phase is bound between 70~K and 118~K within the  pressure range of 200-400~bar. Our simulations yielded somewhat accurate temperature windows for the stability of these phases. However, phase transitions were observed at significantly higher pressures than those reported experimentally. Consequently, we predict the triple point of methane at 70~K and 200~bar, whereas the experimentally obtained  triple point is at 90~K and 0.117~bar \cite{Goodwin1972, Ranieri2024-tg}. This shift in the pressure range results in a broader co-existence of solid-gas phases. Nonetheless, while the ability of the CH GAP to capture the solid-liquid-gas phase transitions exceeded our initial expectations, its inability to predict the phase diagram quantitatively arises from the exclusion of quantum nuclear effects and incomplete consideration of vdW interactions. Due to the light mass of hydrogen atoms, the properties of methane are significantly influenced by quantum nuclear effects, particularly in the liquid phase. As previously discussed, the TS-vdW correction scheme does not fully describe non-bonded interactions; within the weakly bonded regime, even a slight change in the vdW contribution can lead to massive alterations in the phase diagram, e.g., due to inaccurate mass density predictions. In addition, MLPs inherit the limitations of the used exchange-correlation functionals, which result in functional-driven error. Weakly bonded interactions are poorly described even within the DFT framework, where the use of different functionals (and vdW correction schemes) leads to significant discrepancies in the properties. An example of these discrepancies was highlighted in the study by Gillan \cite{Gillan2014} , where the density of water was simulated using different vdW and exchange-correlation functionals, resulting in incorrect density estimations within PBE approximations even after applying many-body dispersion corrections. Given that our CH GAP relies on a dataset generated using the PBE functional, achieving quantitative agreement for these properties is unattainable. Considering the aforementioned points, our CH GAP demonstrates a remarkable level of robustness and general degree of applicability.

We note here that MLPs can be trained to achieve a quantitative description of specific systems at the cost of sacrificing their generality. This has indeed been previously done for methane~\cite{veit_2019}. Thus, in this section we have not introduced our CH GAP as a tool of choice to carry out detailed characterization of the thermodynamic properties of molecular gases or liquids. Rather, we show that it can retain reasonable behavior for these systems while being simultaneously able to describe the thermochemistry of a wide range of hydrocarbons and hydrogenated a-C, thus highlighting its usefulness as an exploratory tool to uncover trends that can be later refined with more purpose-specific MLPs.

\subsection{C-H compounds under extreme conditions}
\begin{figure*}[t!]
\centering
\includegraphics[width=\textwidth]{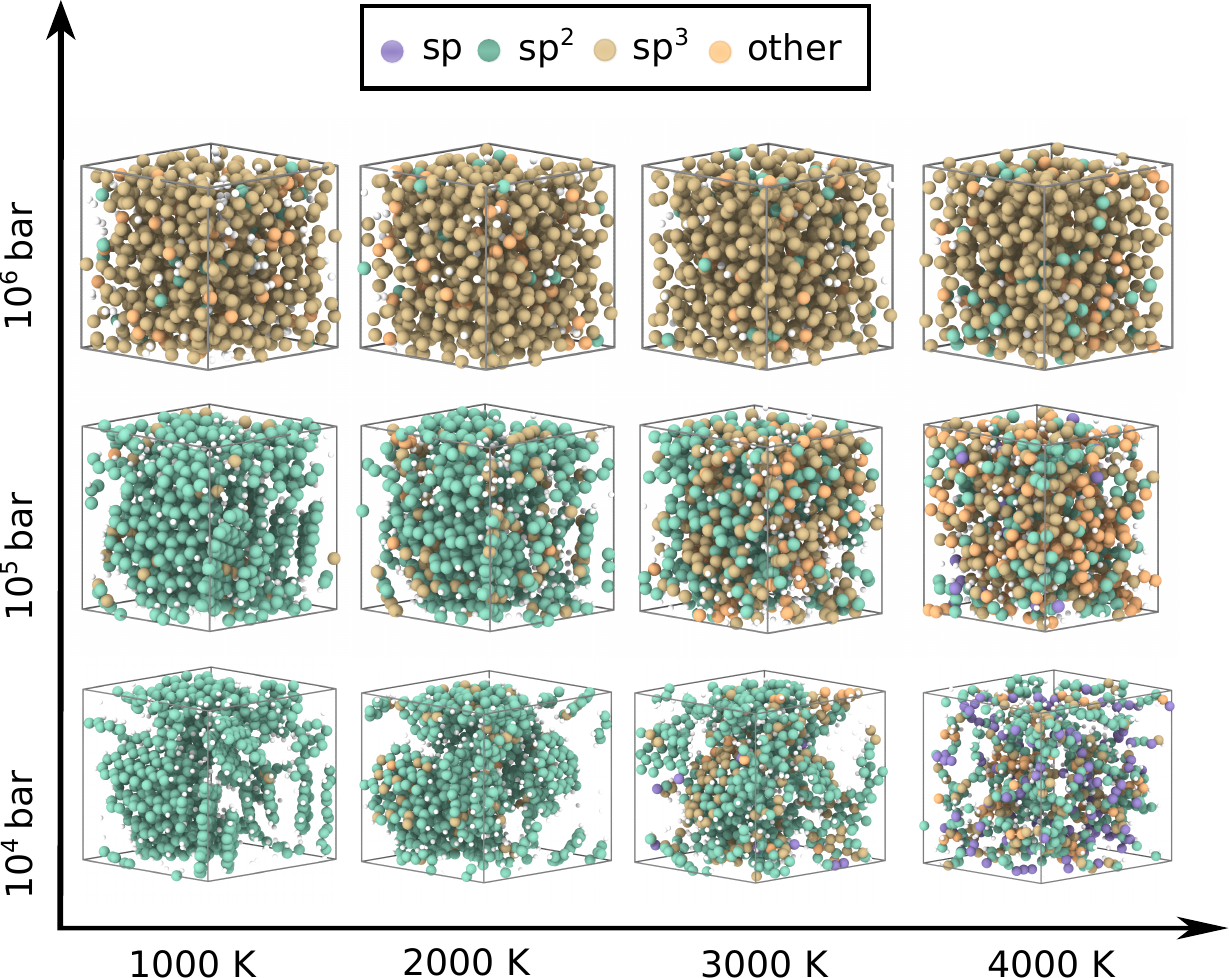}
\caption{\label{fig:smash} Metastable phase diagram for the C-H system from a mixture of coronene and circumcoronene at high pressure (from 10 kbar). The coloring of C atoms is according to the local atomic environment. sp motifs are colored violet, sp$^2$ are green, sp$^3$ are golden, and other motifs are orange.}
\end{figure*}

At the opposite extreme are the simulations of condensed C-H systems subjected to high pressure and temperature (collectively, ``extreme conditions''). Our focus here resides in demonstrating the ability of our potential to unify the simulation of hydrocarbons and hydrogenated carbon materials.
The selection of appropriate precursors for simulating the synthesis of new carbon-based materials under extreme conditions assumes mainly the C to H ratio. Ideally, the chosen candidates should exhibit a high C to H ratio, unlike conventional hydrocarbons which tend to have high hydrogen concentrations. Hence, we used polycyclic aromatic hydrocarbons (PAHs) such as coronene (C$_{24}$H$_{12}$) and circumcoronene (C$_{54}$H$_{18}$) as precursors. Known for their occurrence in coal and petroleum products \cite{EGBEDINA2022107260}, from industrial processes, in mineral deposits and even the interstellar medium and meteorites \cite{Krishnamurthy1992, Flory1967, Kwok2004}, as well as versatility across various material science applications, PAHs serve as viable precursors for the synthesis of carbon-based materials \cite{Davydov2004}. Notably, their utilization significantly reduces the H to C ratio compared to conventional hydrocarbon precursors.

Besides the conventional applications in materials science, the behavior of carbon-based compounds under extreme pressures of up to 10$^6$~bar is relevant in astrophysics, for instance within the context of investigating exoplanetary interiors, stellar environments, ice giants like Neptune, and carbon-rich white dwarfs~\cite{Madhusudhan_2012, Mashian}. Specific examples where it is crucial for the MLP to remain stable under these conditions are: modeling carbon behavior near the surface or shallow layers of large rocky planets, carbon phases within moderately dense exoplanets~\cite{Madhusudhan_2012}, ultra-high pressure conditions in dense star atmospheres~\cite{Ho2009-ft}, and for understanding conditions similar to those within Earth's mantle and outer core regions~\cite{fischer2020}.

To simulate the synthesis of CH-based materials under extreme conditions, an equal mixture of coronene and circumcoronene was prepared and then equilibrated for 100~ps at 1000~K for each value of pressure. 
The temperature was controlled using the Bussi thermostat with a time constant of 100~fs and a time step of 0.5~fs. 
For pressure coupling, the Berendsen barostat with a time constant of 1~ps was applied. Then, the structures were equilibrated for an additional 20~ps, and the average box dimensions over the last equilibration were obtained. In the final step, the barostat was turned off and a box scaling transformation was used to slowly settle the box dimensions of the structures to the average values obtained earlier. After this step, the last snapshot was used as a starting configuration for the next value of temperature. Using this procedure, we sampled different temperatures at various pressures and obtained the metastable phase diagram shown in Fig~\ref{fig:smash}. The atoms are colored according to their sp, sp$^2$, and sp$^3$ structural motifs. Under a pressure of 10$^4$~bar and a temperature of 1000~K, all molecules remain intact, forming a compact, layered material with intermolecular distances of approximately 3.5--3.7~\AA{}. These structures predominantly exhibit sp$^2$ motifs, similar to the precursor PAH molecules, up to 3000~K. At 3000~K, partial molecular breakdown and the formation of sp$^3$ bonds are observed. At 4000~K, extensive molecular decomposition occurs, resulting in a mixture of sp, sp$^2$, and sp$^3$ bonding. The density of these structures shows slight variation with temperature but remains approximately 1.1 g/cm$^3$.

Under a pressure of 10$^5$~bar and a temperature of 1000~K, most molecules remain intact, forming closely packed, compressed layers of coronene and circumcoronene molecules with various orientations. Consequently, sp$^2$ motifs predominate with the intermolecular distance decreasing to approximately 2.8--3.0 \AA{}. As the temperature increases, there is a notable rise in the prevalence of sp$^3$ motifs. At 2000~K, the sp$^3$ content reaches 16~\%; at 3000~K, it increases to 43~\%; and at 4000~K, it reaches 54~\%. Simultaneously, the density of these structures increases, reaching approximately 2~g/cm$^3$.

Under a pressure of 10$^6$~bar, significant structural changes occur, with all precursor molecules decomposing and H$_2$ gas forming. Consequently, additional steps were taken to remove H$_2$ molecules throughout the simulations. As a result, all structures at this pressure exhibit predominantly sp$^3$ bonding, forming a dense, amorphous carbon structure with remaining hydrogen localized in small pockets. C--C bond lengths shorten to 1.4 \AA{} and the density increases significantly to 3.8 g/cm$^3$, similar to the density of diamond (3.5 g/cm$^3$). After slowly releasing the high pressure to 1 bar and equilibrating, the density decreases to 3.0 g/cm$^3$.

\begin{figure}[t]
\centering
\includegraphics[width=\columnwidth]{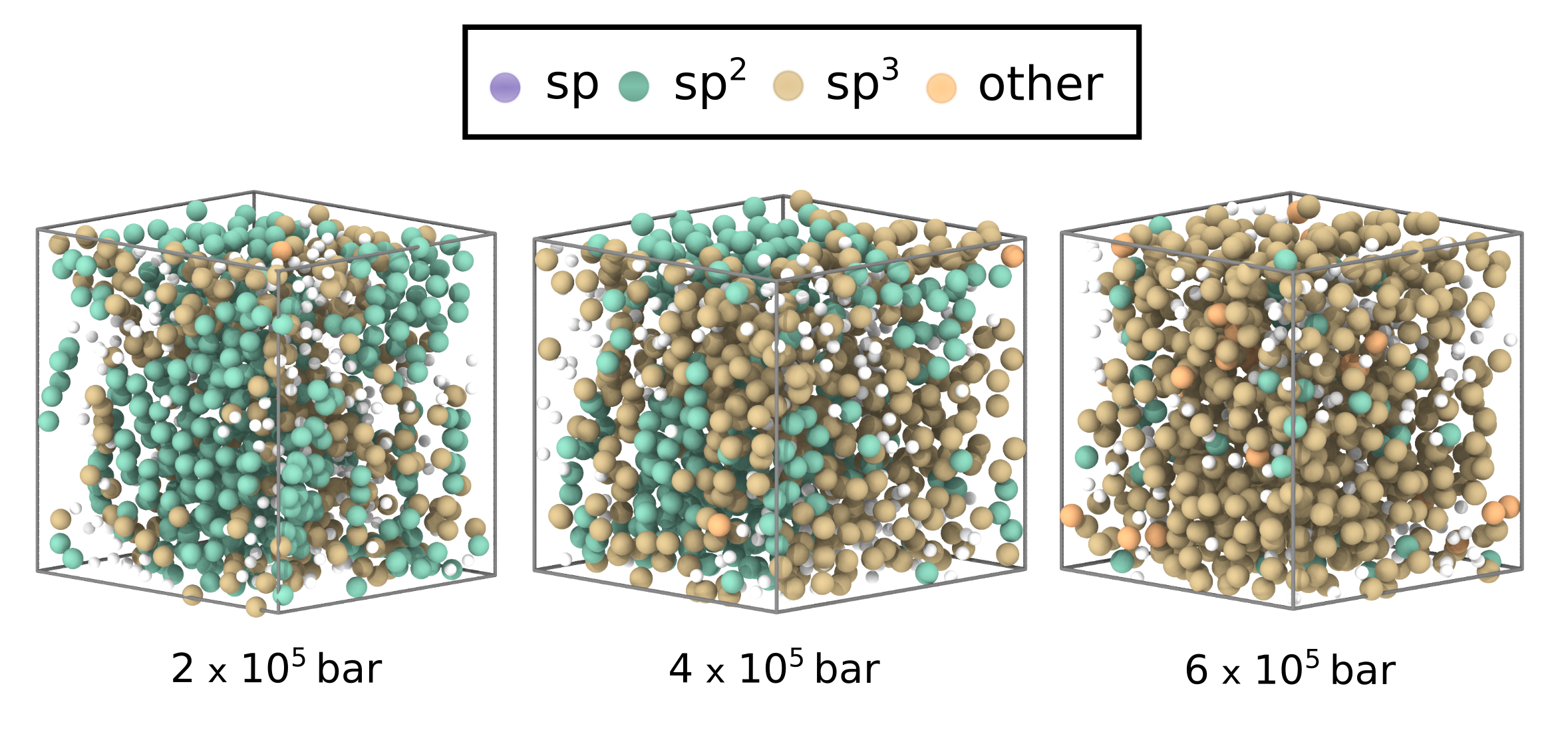}
\caption{\label{fig:105} Transformation of C-H structures at the temperature of 2000~K and the set of pressures of 2, 4, and 6$\times$10$^5$ bar.}
\end{figure}

We observe that the primary transformation from sp$^2$-dominant to sp$^3$-dominant structural motifs occurs within the 10$^5$ to 10$^6$~bar pressure range. Thus, we sampled this range of pressures to detect the gradual structural transformations. The C-H structures at 2000~K under pressures of 2, 4, and 6$\times$10$^5$~bar are shown in Fig.~\ref{fig:105}. Under a pressure of 2$\times$10$^5$~bar,  the sp$^3$ content increases to 36~\%, much more formed H$_2$ molecules are removed throughout the simulation. At the pressure of 4$\times$10$^5$~bar the sp$^3$ content reaches 58~\% and increases to 92~\% at 6$\times$10$^5$~bar.

Available experimental data on the high-pressure modification of PAH molecules are sparse. Specifically, data on the modification of coronene compounds are particularly limited. Additionally, reported structural transformations occur under various ranges of pressure and temperature conditions. This variability is largely attributed to differences in high-pressure setups, experimental characterization techniques and their respective accuracy. The variations in experimental setups, including the materials used for sample holders, contribute to this variability. For instance, graphite-based holders  may act as catalysts in the formation of diamond derivatives, potentially leading to lower temperature and pressure requirements.  Despite these challenges, we can identify the common trends across these studies.

Earlier studies of coronene compounds consistently report the formation of dimers and stacked molecular clusters \cite{calvo2005, HRojas2019}. Our simulations also observe these stacked clusters, aligning with findings from Ref.~\cite{oligomer}, where such formations occur at a pressure of 35 kbar. The oligomerization then leads to the graphitization process, which is mostly controlled by changes in temperature (T \(>\) 873 K) in these experiments. In our simulations, graphitization is detected between 10 and 100 kbar and controlled primarily by a pressure change. Due to the constraints of molecular dynamics simulations, we employ higher temperatures to accelerate rare events within the limited simulation timeframe. Thus, the temperature ranges in our simulations cannot be expected to fully match the experimental conditions quantitatively.  
Experimental observations of structural transformations from graphitic to diamond-like structures have been reported at pressures of 1.2 $\times$ 10$^5$ bar \cite{onodera2000synthesis} and 8 $\times$ 10$^5$ bar \cite{Davydov2004}. Our findings are consistent with these conditions, indicating that dehydrogenation and structural rearrangement commence at pressures as low as 1 $\times$ 10$^5$ bar at high temperatures (3000 K), with these transformations becoming more pronounced at higher pressures and even at lower temperatures. As illustrated in Fig. \ref{fig:105}, the rapid increase of sp$^3$ content is accompanied by the formation of H$_2$ gas, as observed in experimental studies \cite{Davydov2004}. Additionally, we observe that the diffusion of hydrogen occurs on the edges of misaligned stacked clusters, while increasing the pressure results in the formation of pockets with hydrogen accumulating within them. To mimic experimental conditions, where degassing is observed~\cite{Davydov2004}, we removed the H$_2$ molecules and continued applying pressure. 

The CH GAP potential demonstrates good agreement with the available experimental data and predicts structural changes under high-pressure conditions in semi-quantitative agreement with experiments. We conclude that the CH GAP is a useful tool for simulating the synthesis of carbon-hydrogen based materials under these conditions, and could potentially be used for understanding carbon's role in galactic chemical evolution.

\section{Conclusions}
In summary, we have developed a comprehensive structural database for carbon- and hydrogen-containing systems and trained an accurate general-purpose GAP interatomic potential from it. Both the general-purpose GAP and training database are publicly available and the model can be retrained for a particular need. Our developed CH GAP unifies the description of different hydrocarbons and hydrogen-containing carbon materials, and demonstrates exceptional predictive capabilities across a broad spectrum of structures and properties pertinent to carbon- and hydrogen-based structures under diverse thermodynamic conditions.
Our results highlight the predictive power of the CH GAP potential and its ability to explore complex hydrocarbon chemistry, including the formation of various alkanes, aromatics and their networks, through simple modifications in temperature and the carbon-to-hydrogen ratio. The potential retains reasonable behavior for weakly bonded systems, including interacting molecules and methane at low temperatures while being simultaneously able to describe the thermochemistry of a wide range of hydrocarbons and hydrogenated a-C.

Furthermore, simulations of C-H material formation under extreme conditions underscore the versatility of our CH interatomic potential in modeling materials in challenging environments. The general-purpose CH GAP presented here holds significant potential for impact in various fields, including organic chemistry, materials science, energy storage and conversion, and astrochemistry. Particularly, in astrochemistry, where direct access to target compounds is often limited, computational simulations are particularly valuable and can facilitate future research in this domain.

\begin{acknowledgments}
The authors acknowledge financial support for this project by the Research Council of Finland under projects 330488, 336304, 358050 and 355301, as well as computational resources from CSC (the Finnish IT Center for Science) and the Aalto University Science-IT project.
\end{acknowledgments}

\end{document}